\documentclass[fleqn,usenatbib]{mnras}

\usepackage{newtxtext,newtxmath}

\usepackage[T1]{fontenc}
\usepackage{ae,aecompl}

\usepackage{graphicx}	
\usepackage{amsmath}	
\usepackage{amssymb}	

\usepackage[disable]{todonotes}   

\usepackage{xcolor}

\newcommand\hl[1]{#1}

\newcommand{\for}{\text{for }}   

\newcommand*\dif{\mathop{}\!\mathrm{d}}

\title[The ground layer of turbulence at Paranal]{Characterisation of the ground layer of turbulence at Paranal using a robotic SLODAR system}

\author[T. Butterley et al.]{
T. Butterley,$^{1}$\thanks{E-mail: timothy.butterley@durham.ac.uk (TB)}
R. W. Wilson,$^{1}$
M. Sarazin,$^{2}$
C. M. Dubbeldam,$^{1}$
J. Osborn$^{1}$
\newauthor  and P. Clark$^{1}$\\
$^{1}$Centre for Advanced Instrumentation, Department of Physics, University of Durham, South Road, Durham, DH1 3LE, UK\\
$^{2}$European Southern Observatory (ESO), Karl-Schwarzschild-Str. 2, D-85748 Garching, Germany\\
}

\date{Accepted XXX. Received YYY; in original form ZZZ}

\pubyear{2019}

\begin{document}
\label{firstpage}
\pagerange{\pageref{firstpage}--\pageref{lastpage}}
\maketitle


\begin{abstract}
We describe the implementation of a robotic SLODAR instrument at the Cerro Paranal observatory. The instrument measures the vertical profile of the optical atmospheric turbulence strength, in 8 resolution elements, to a maximum altitude ranging between 100~m and 500~m. We present statistical results of measurements of the turbulence profile on a total of 875 nights between 2014 and 2018.  The vertical profile of the ground layer of turbulence is very varied, but in the median case most of the turbulence strength in the ground layer is concentrated within the first 50~m altitude, with relatively weak turbulence at higher altitudes up to 500~m. We find good agreement between measurements of the seeing angle from the SLODAR and from the Paranal DIMM seeing monitor, and also for seeing values extracted from the Shack--Hartmann active optics sensor of VLT UT1, adjusting for the height of each instrument above ground level. The SLODAR data suggest that a median improvement in the seeing angle from 0.689~arcsec to 0.481~arcsec at wavelength 500~nm would be obtained by fully correcting the ground--layer turbulence between the height of the UTs (taken as 10~m) and altitude 500~m. 
\end{abstract}


\section{Introduction}


The ground layer of \hl{atmospheric} optical turbulence, located within a few hundred metres of the surface, 
typically contributes a substantial fraction of the total atmospheric turbulence strength \citep{Tokovinin03,Chun09}.
Hence ground layer adaptive optics (GLAO) 
systems have been developed to correct only the low altitude turbulence. 
For low altitude turbulence the isoplanatic field of view 
for adaptive optics (AO) 
correction is large, so that partial image correction can be effected over a large field 
of view. The degree of correction achievable with GLAO is determined by the fraction of the 
total turbulence to be found in the ground layer, above the height of the telescope. The field of view for effective GLAO correction 
depends on the vertical distribution of the ground layer above the telescope \citep{Rigaut02,Tokovinin04}.


Statistical measurements of the vertical distribution of turbulence close to the ground are 
therefore of interest in modelling the performance of proposed and existing GLAO systems. 
Real-time turbulence measurements can be used to optimise the running parameters of such 
systems and to monitor whether the optimum level of image correction is being delivered, 
given the current atmospheric conditions.  In the case where there are significant time 
overheads involved in starting an AO observation a real-time measure of the 
fraction of ground layer 
turbulence can be used to determine whether conditions are favourable for GLAO.

The Adaptive Optics Facility (AOF) 
\citep{Kuntschner12,Madec18} \todo{other AOF refs?} at Paranal observatory is an upgrade to one of the 8~metre Unit Telescopes (UTs) 
to include an adaptive secondary mirror, 4 laser guide stars (LGS) 
and 2 AO modules: GRAAL and GALACSI. GRAAL is a ground layer AO module for the Hawk-I infrared wide-field imager, with a science field of $7.5' \times 7.5'$. GALACSI increases the performance of the Multi-Unit Spectroscopic Explorer (MUSE) 
instrument with two AO modes: in wide field mode GALACSI delivers ground layer AO correction with a $1' \times 1'$ field of view and in narrow field mode it delivers tomographic AO correction with a $7.5'' \times 7.5''$ field of view. 
To predict AOF performance in GLAO mode requires information on the ground layer turbulence profile up to approximately 500~m.

Strong turbulence often occurs within a few tens of metres of the ground, where the surface 
wind interacts directly with the ground and local topography, and the air may be heated 
(or cooled) strongly by the ground. The largest astronomical telescopes may be taller  
than the typical scale height of this surface layer of turbulence. They may then experience 
significantly better seeing conditions than a smaller telescope on the same site. 
The detailed structure of the turbulence profile within the first 50~m altitude can 
therefore give an improved understanding of the observing conditions for the different 
telescopes and instruments at a site \citep{Sarazin08}.

For the analysis and discussion presented here, it is helpful to define this surface layer 
turbulence contribution as a distinct component of the ground layer. Hence here we define the 
surface layer to refer to turbulence at altitudes below 50~m, with the ground layer extending 
to 500~m altitude.


The importance of the ground layer turbulence has been recognised in the development of a number of instruments and monitors specifically to measure it, and exploited for characterisation of the major observatory sites and in site selection campaigns for the next generations of extremely large telescopes. These include: sonic detection and ranging (SODAR) \citep{Els09}, low layer SCIDAR (LOLAS) \citep{Avila08}, the lunar scintillometer (LuSci) \citep{Tokovinin10, Hickson13, Lombardi13} and mast-mounted sonic anemometers \citep{Aristidi18}. \hl{A multi-instrument study of the surface layer at Paranal was made by \citet{Lombardi10}.}

\begin{figure}   
    \includegraphics[width=\columnwidth]{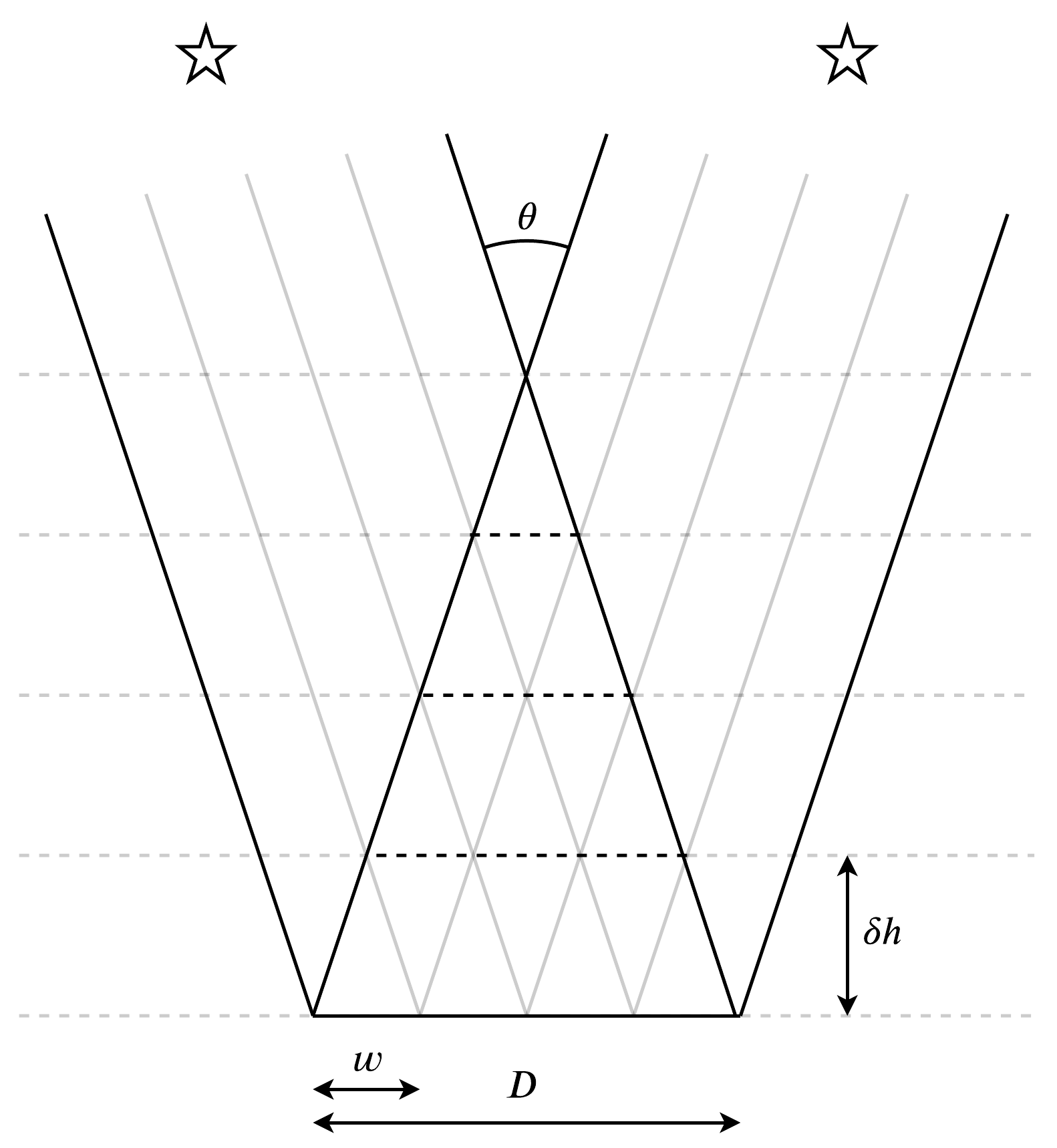}
    \caption{Overview of the SLODAR method geometry. $D$ is the telescope aperture diameter, $w$ is the subaperture width and $\theta$ is the separation of the target stars. The size of a vertical resolution element (for a target at zenith) is $\delta h = w/\theta$ .}
    \label{fig:geom}
\end{figure}

The slope detection and ranging (SLODAR) method \citep{Wilson02, Butterley06b} was developed in the context of the 
Very Large Telescope (VLT)/Extremely Large Telescope (ELT) 
and was first deployed at the Paranal observatory in 2005 \citep{Wilson09}. SLODAR is an 
optical `crossed-beams' method in which the optical turbulence profile is recovered from 
the cross-covariance of Shack-Hartmann wavefront sensor (WFS) measurements of the wavefront 
phase gradient for a pair of stars with known angular separation. The vertical resolution 
of the technique improves as the angular separation of the target stars increases, but 
with a consequent reduction in the maximum altitude to which direct measurements extend, \hl{as illustrated in figure~\ref{fig:geom}.}
The total number of resolution elements is fixed, and is equal to the number of 
sub-apertures of the wavefront sensor subtended across the telescope aperture. In its 
original format Paranal SLODAR, based on a 0.4~m telescope, exploited target stars with a separation 
of $\sim$1~arcmin, to provide an eight point profile reaching a maximum altitude of approx. 1~km. 

A later development allowed for the use of target stars with much larger separations, 
$\sim$~5~--~15~arcmin. For these large separations, separate WFS optics and 
detectors are used for each target star, since they could not be imaged directly onto a 
single detector. In this format, known as surface layer SLODAR (SL-SLODAR), 
\citep{Osborn10} a vertical 
resolution of less than 10~m can be achieved. The instrument can then resolve the structure 
of the optical turbulence profile on scales substantially smaller than the height of the 
telescope structures at the Paranal site (the domes of the unit telescopes of the VLT are 30~m high). 

\begin{figure}   
    \includegraphics[width=\columnwidth]{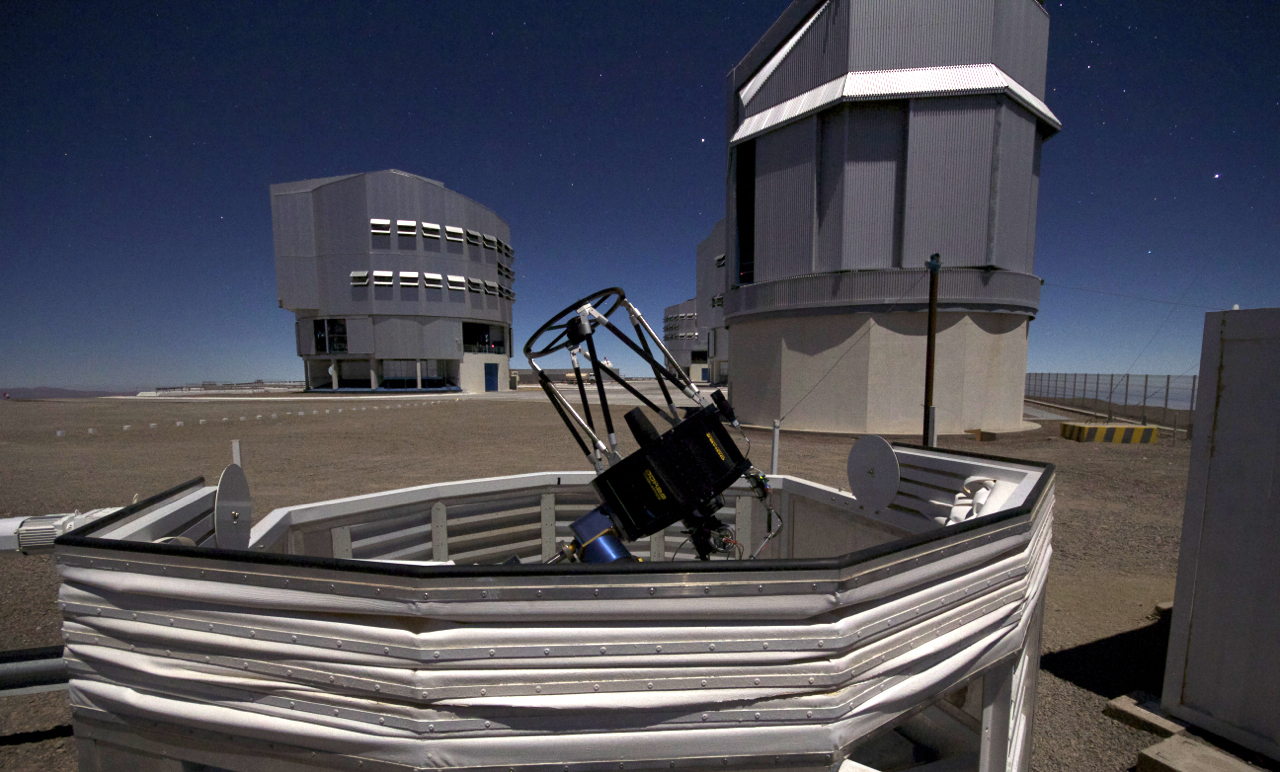}
    \caption{Photograph of the Paranal robotic SL-SLODAR instrument (foreground) with the VST (on the right) and UTs (left and behind the VST) in the background.}
    \label{fig:photo}
\end{figure}

The SL-SLODAR has been developed into a fully robotic system (shown in figure~\ref{fig:photo}) by Durham University in collaboration with the European Southern Observatory (ESO). It was installed at Paranal in 2013 and commissioning by Durham University was completed by mid-2014. Since then the instrument has been integrated into the astronomical site monitor (ASM), a suite of instruments that constantly monitors the ambient conditions at the observatory site. The SL-SLODAR provides surface layer and ground layer profiling to support the AOF.

This paper is organised as follows. In section~\ref{sec:instrument} we describe the robotic SL-SLODAR system at Paranal including hardware, software and data analysis methods.
In section~\ref{sec:beta} we discuss limitations due to poor convergence in low wind speeds.
In section~\ref{sec:stats} we present results from the first years of observations. These include statistics of the strength and vertical profile of the ground--layer of optical turbulence above the site, relevant to GLAO correction and also to the 
seeing angle as a function of height above surface level \hl{(for uncorrected images i.e. for seeing limited observations through a telescope above the ground)}.
We also present cross- comparisons of the data with other seeing monitors and turbulence profilers 
operating at the site, including: a differential image motion monitor (DIMM); a multi-aperture scintillation sensor (MASS); 
image full width at half maximum (FWHM) measurements 
from the Shack-Hartmann sensors of the active optical systems of the UTs of the VLT itself. 
Section~\ref{sec:conc} contains our conclusions.








\section{Instrument description}
\label{sec:instrument}

\begin{figure}   
    \includegraphics[width=\columnwidth]{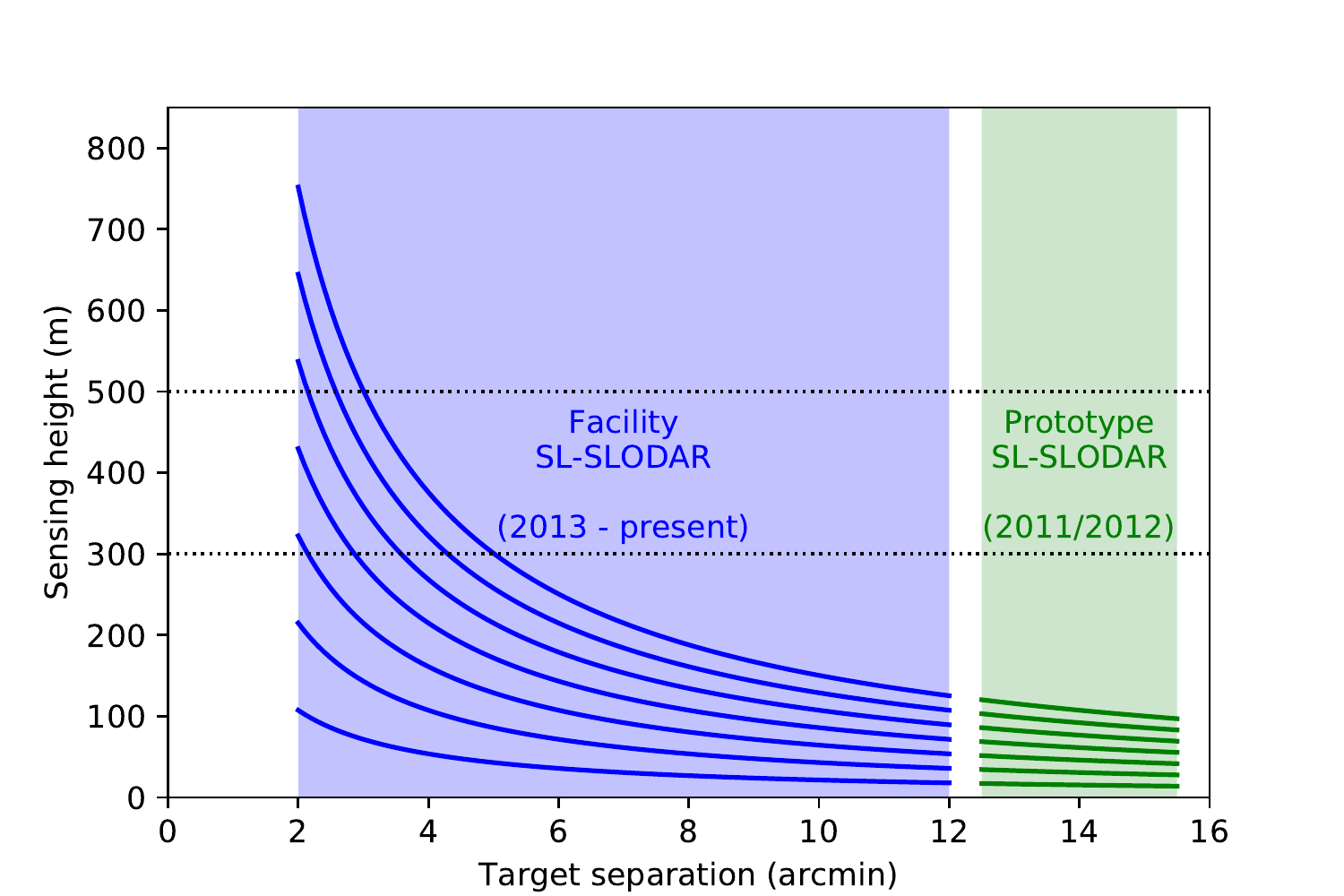}
    \caption{Heights of the 8 fitted SLODAR layers as a function of target star separation for a target at zenith. The first layer is always at 0~m. The blue and green regions show the range supported by the current facility instrument and the earlier prototype instrument respectively.}
    \label{fig:resolution}
\end{figure}

The SL-SLODAR instrument  
consists of a 0.5~m telescope equipped with a pair of $8~\times~8$ subaperture Shack-Hartmann wavefront sensors that can observe stars with separations ranging from 2~arcmin to 12~arcmin. The turbulence profile is recovered from the spatial cross-covariance of wavefront slope measurements from the two stars. The instrument delivers 8-layer profiles of the ground layer of turbulence. The vertical resolution and maximum sensing altitude depend on the separation of the target stars (see figure~\ref{fig:resolution}) and the zenith angle of observation; the maximum possible sensing height is approximately 500~m. 

While the instrument supports the full range of star separations between 2 and 12 arcmin, in practice it is generally desirable to observe targets at the extremes of this range. The narrow target regime (2--5 arcmin) is used to profile the ground layer up to 500~m or as close to 500~m as possible.  The wide target regime (10--12 arcmin) is used to measure the surface layer with the best resolution possible.

Much of the time, especially when the instrument is observing the narrow targets required to reach a maximum altitude of 300-500~m, the surface layer of turbulence is too thin to be resolved. The surface layer is therefore usually observed entirely in the first resolution element and the instrument is unable to determine what fraction of the surface layer turbulence is observed by the UTs.

Prior to commissioning of the facility SL-SLODAR, a prototype version of the instrument was operated (2011--2012). The prototype used even wider separation targets (typically 13--15 arcmin as shown in figure~\ref{fig:resolution}) and was able to resolve the surface layer. This period is therefore a source of statistical information that can be used to construct an average model of the surface layer. This model, scaled by the total turbulence strength in the first resolution element, can be used as our best estimate of the surface layer profile in data where the resolution is insufficient to resolve the surface layer. 

\subsection{Optomechanical design}
\label{sec:instDesign}

\begin{figure}    
	\includegraphics[width=\columnwidth]{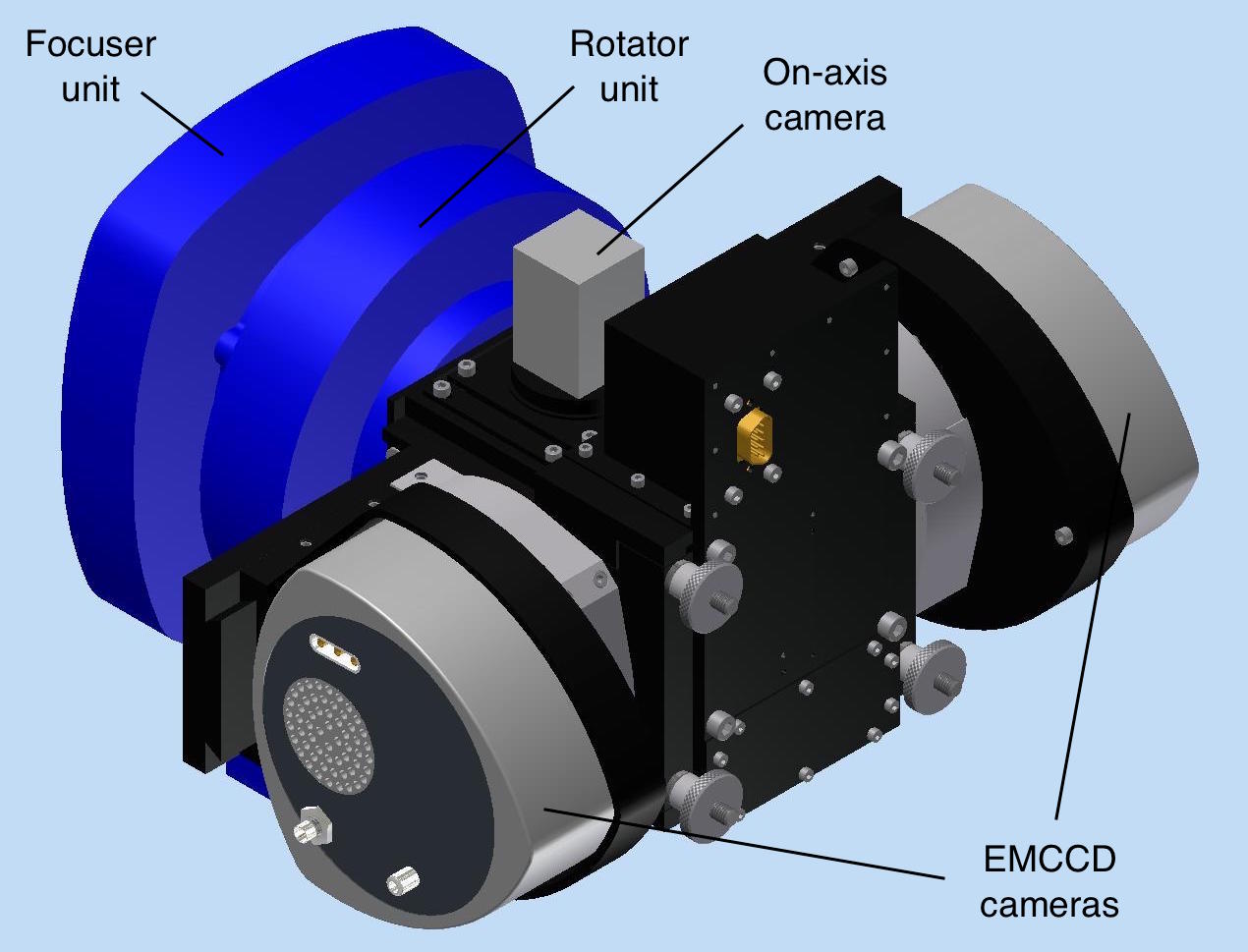}
	\caption{CAD image of the fully assembled SL-SLODAR instrument (excluding cables).}
	\label{fig:instBackend}
\end{figure}

\begin{figure}    
	\includegraphics[width=\columnwidth]{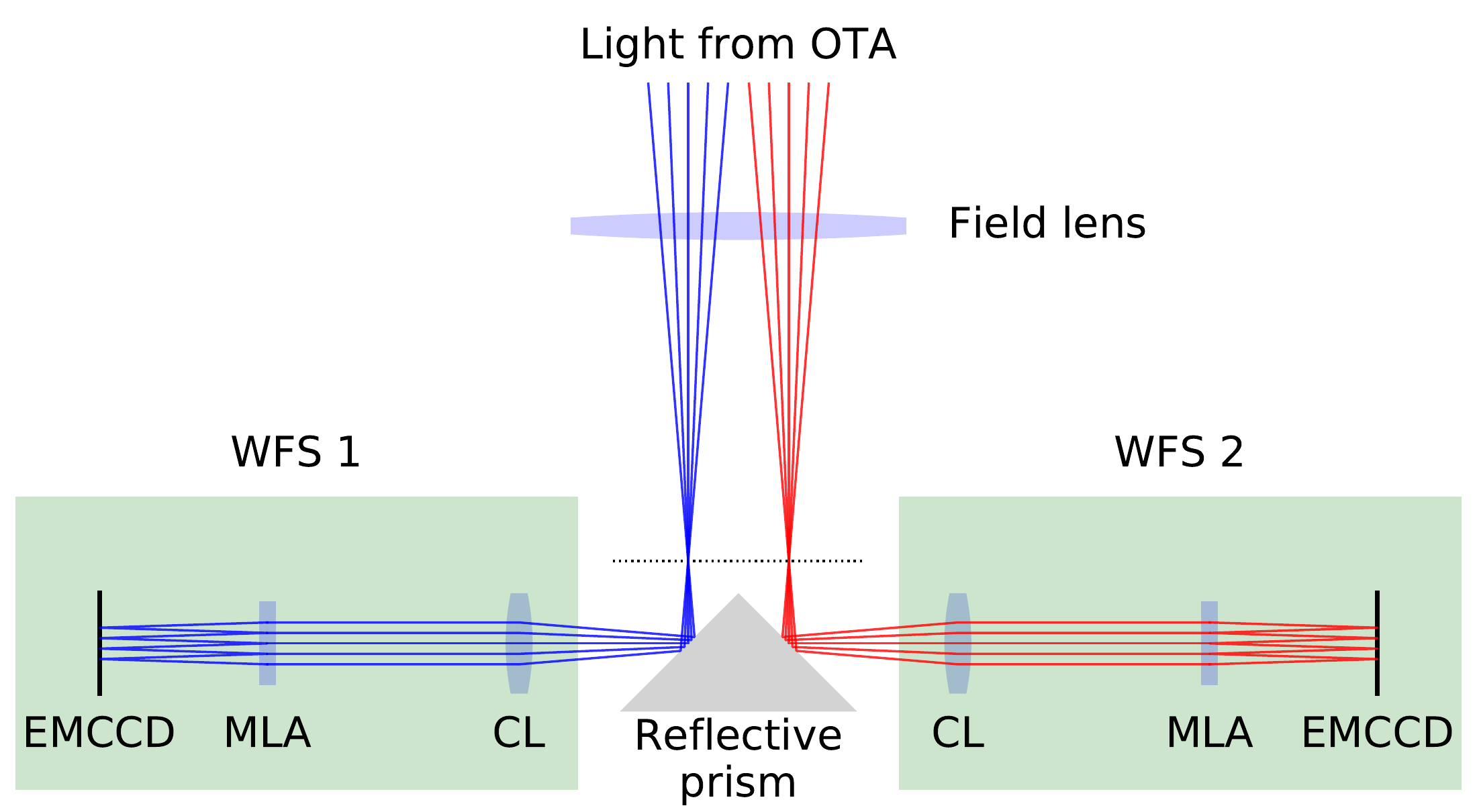}
	\caption{Optical layout of the SL-SLODAR instrument. CL and MLA denote collimating lens and microlens array respectively. Light from two different stars is shown as red and blue rays. The dotted line shows the location of the focal plane of the telescope.}
	\label{fig:opticalLayout}
\end{figure}

\begin{figure*}    
    \centering
	\includegraphics[width=2\columnwidth]{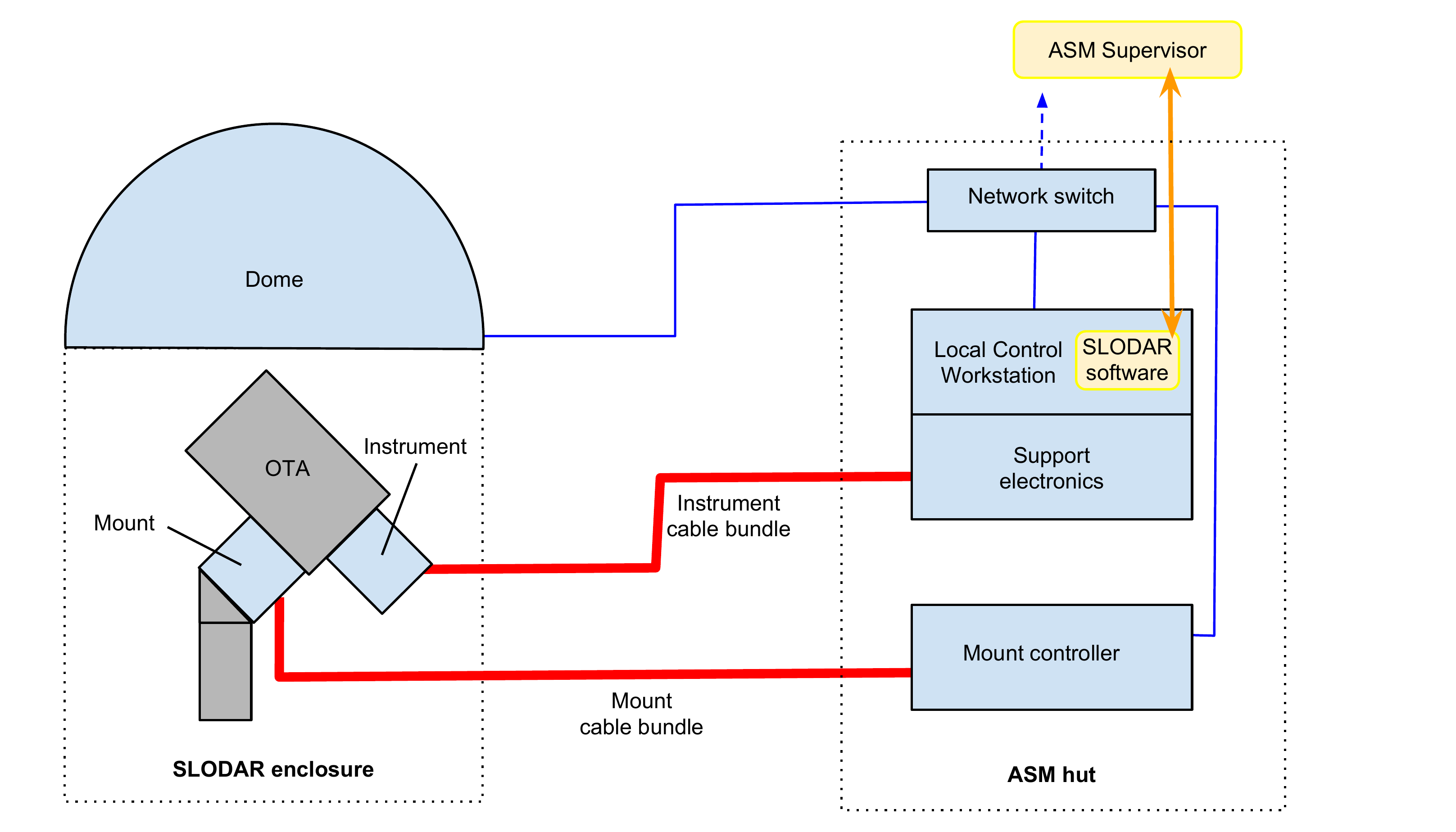}
	\caption{Robotic SL-SLODAR system overview. Blue boxes represent electrical/electronic devices. Yellow boxes represent software. Blue lines are network connections. Red lines are custom cable bundles. The orange line represents network communication between software components.}
	\label{fig:overview}
\end{figure*}

The robotic SL-SLODAR system is based on a 0.5~m optimized Dall-Kirkham reflecting telescope on an Astelco NTM500 German equatorial mount. The SLODAR wavefront sensing instrument (figure~\ref{fig:instBackend}) is installed at the Cassegrain focus. The design of the instrument requires the focal plane of the telescope to be telecentric; this is achieved by the inclusion of a field lens, which has a focal length of 1180~mm and is mounted to the optical tube assembly (OTA) such that it is positioned approximately 40~mm before the telescope focus.

The instrument is attached to the OTA via mechanical rotator and focuser units which allow the entire instrument to be rotated and translated longitudinally.

Light from two stars enters the instrument and encounters a reflective  prism  close  to  the  telescope focus, as shown in figure~\ref{fig:opticalLayout}. The rotator is set to align the wavefront sensor arms in the same orientation as the vector between the two stars.  The light  from  the  two  stars  is  reflected,  in  opposite  directions,  into  the two WFS assemblies.  The  prism is mounted on a linear stage that positions it  along  the  optical  axis  of  the  telescope,  depending  on  the  separation  of  the  stars,  such that the reflected beams enter the WFS assemblies through the  centres  of  their  collimating  lenses.  The  focuser  is  set  such  that  the  total  path  length of the light does not change as a result of moving the prism. 


The focuser has a travel of approximately 9~mm. This is  the  factor  that  limits  the  maximum  target  separation  that can be accommodated by the  instrument. The minimum separation is that required to avoid the beams vignetting on the point of the prism.

Each WFS arm comprises a collimating lens and microlens array (MLA) that images the spot pattern directly onto a detector. 
The detectors are Peltier/air cooled, $640 \times 480$ pixel electron multiplication CCD cameras (EMCCD), model Andor~Luca~S. During normal SLODAR operation, the cameras  operate with an exposure time of 3~ms 
and a frame rate of 57.6~Hz. 

The instrument includes a further mechanism that can introduce a 45$^\circ$ pick-off mirror (also on a linear stage) into the beam before it reaches the prism and direct it to an on-axis camera on top of the mounting block. The telescope focus lies several mm in front of this detector so the star image formed on the  detector is defocused. This `calibration  mode' allows  a  single  on-axis  star  to be  observed, and is used to update the pointing model for the mount. In addition to the pick-off mirror, the linear stage also carries a 1~mm  wide  slit  mask that is aligned  along  the  axis  of  the  WFSs. During normal operation this slit is centred on the optical axis to reject scattered moonlight, sky background and unwanted stars from the WFSs.

\todo{JO: Environment?}

The SLODAR instrument is located at the north-eastern edge of the VLT observing platform, approximately 100~m northeast of UT4. 
The telescope is contained within an automated enclosure 
that protects it from the elements when it is not operating (visible in figure~\ref{fig:photo}). The enclosure (also referred to as the `dome') has a retractable canvas hood enclosure. The sides are louvred to permit air flow through the enclosure; this is to prevent warm air getting trapped inside the enclosure and generating local turbulence. Completely open sides would allow better air flow but would offer no protection against rain, snow or dust contamination.

The control electronics are contained in the `ASM hut', a service building a few metres away from the dome. These consist of a `local control workstation' (LCW) running Scientific Linux 6.4, 
the telescope mount control computer, power supplies and controllers for the instrument mechanisms, and a network power controller. 
The cameras have a USB interface so powered USB extenders are required to cover the 12 metre distance between the telescope and the LCW and other electronics in the ASM hut.

\subsection{Alignment}

The WFS module optics were initially aligned off-sky using a telescope simulator, which simulates stars at a range of off-axis angles.  
The  stars  are  simulated  by  imaging  the  ends  of  a  row  of  optical  fibres  through  a  simple  2-lens  telecentric  optical  system  with  1:1  magnification.  The  image  plane  of  this  system  matches  the  characteristics  of  the  telescope focal plane to a good approximation. The separation of the collimating lens and MLA is set by examining the WFS spot illumination pattern and ensuring it is the same for all illumination angles. This ensures the MLA is conjugated to the pupil of the telescope.

After aligning the WFS modules using the simulator, a final adjustment must be made on sky: the transverse position of each MLA must be set to produce a symmetrical Shack-Hartmann spot pattern.

\subsection{Target catalogue}
\label{sec:targCat}

The SL-SLODAR target catalogue was compiled from the Tycho-2 star catalogue.\footnote{\url{http://www.astro.ku.dk/~erik/Tycho-2/}}
Suitable targets were identified by searching the catalogue for pairs of stars that meet the following criteria:
\begin{enumerate}
\item Separation in the range 2 to 12 arcmin.
\item Declination in the range $-70^\circ$ to $10^\circ$.
\item V-band magnitude brighter than 6.5 (for each star in the pair).
\item No other stars in the Tycho-2 catalogue (which is complete to V$\sim$11) within 4 arcmin of either star in the pair. 
\end{enumerate}

\begin{figure}
	\includegraphics[width=\columnwidth]{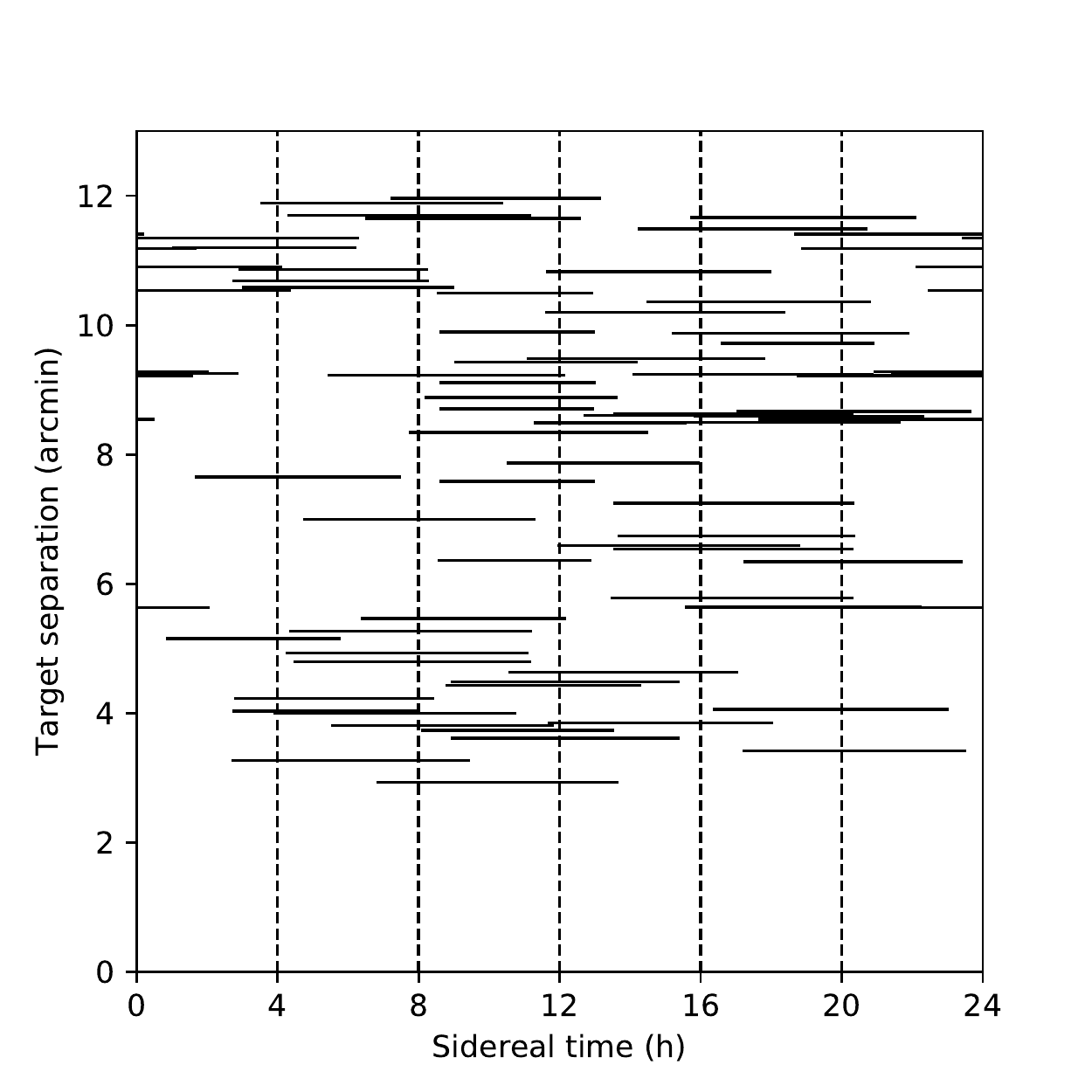}
	\caption{Target availability as a function of local sidereal time. Each horizontal line indicates the LST range for which a particular target is available, with the position on the y-axis indicating the separation of the target.}
	\label{fig:targetSiderealTime}
\end{figure}

\begin{figure}
	\includegraphics[width=\columnwidth]{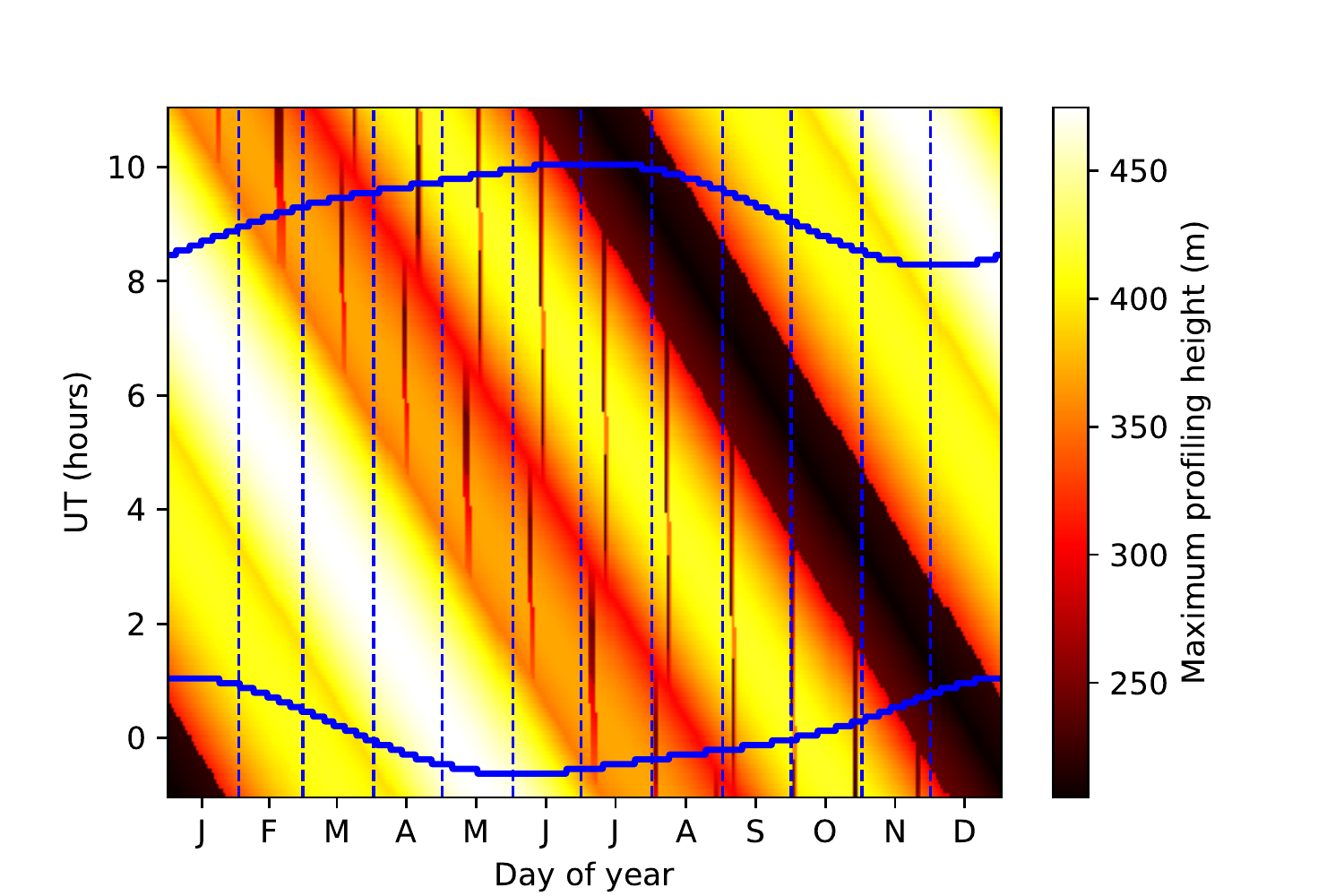}
	\caption{Maximum profiling height available during the course of 2014. Blue traces indicate sunrise and sunset. The near-vertical dark streaks represent times when the moon is close to the narrowest target so a wider target must be used. The very dark diagonal band corresponds to the period around LST~$=1$ hour when the narrowest targets available have separations of $>5$~arcmin.}
	\label{fig:targetMaxHeight}
\end{figure}

Figure~\ref{fig:targetSiderealTime} shows target availability over time -- each horizontal trace shows the period of local sideral time (LST) during which a target is above 45 degrees elevation. There is a period of approximately 4 hours, centred around LST~$= 1$ hour, during which the narrowest target available is wider than 5~arcmin so the instrument can not observe in the low resolution/high maximum sensing altitude regime. The rest of the time there are always at least two targets available with separation $<4$~arcmin. Wide targets are plentiful so it is always possible to observe in the high resolution regime.

Figure~\ref{fig:targetMaxHeight} shows the maximum sensing altitude (i.e the height of the 8th SLODAR resolution element) as a function of time of night and time of year. This accounts for the target elevation and the moon position (for the year 2014). 

\subsection{Observing strategy}


Internal control of the SL-SLODAR system, with the exception of the dome, is handled by a program called the `pilot'. The pilot receives top level commands via a network socket from an external supervisor program. The supervisor controls the dome directly to minimise the risk of a hardware failure preventing it from closing. It is the responsibility of the supervisor to check the ambient conditions; the system does not operate when wind speed is higher than 13~m/s (as the telescope would shake too much) or when the ambient humidity is above 70\% (to prevent dew forming on the exposed optical surfaces). \hl{In these conditions AOF must operate without prior information about the ground layer from the SL-SLODAR system.} When the conditions are good and it is sufficiently dark, the supervisor opens the dome and instructs the pilot to observe. The pilot then initialises all subsystems, slews the telescope to an appropriate target and begins data acquisition.

The pilot  generates  its  current  valid  list  of  targets  at  any  moment  in  time  by  filtering the target catalogue (see section~\ref{sec:targCat}) to exclude targets that are: 
\begin{enumerate}
\item below $45^\circ$ elevation, 
\item less than $15^\circ$ from the moon, or 
\item outside the current specified separation range. 
\end{enumerate}
The valid target list is then sorted by how long each target can be tracked for before it  crosses  the  meridian  or  drops  below  the  elevation  limit.  The  target  that  will  be  valid for longest is at the top of the list. 

The system always maintains a list of at least 3 valid targets to ensure an alternative target is available in the event that a cloud or the Rayleigh plume of a LGS enters the SL-SLODAR WFS field and interrupts data acquisition.  This is achieved by making the 3rd  criterion above flexible when necessary; if filtering the catalogue as described above yields fewer than 3 valid targets then the accepted separation range is widened incrementally until there are at least 3. 

Normally, when the system slews to a new target it chooses the one at the top of the  current  valid  target  list.  Left  to  its  own  devices  it  will  track  this  target,  measuring  profiles  continuously,  until  the  target  becomes  invalid  (by  reaching the meridian  or  falling  too  low in elevation).  The  system  will  then  refresh  the  valid  target  list  and  automatically slew to whichever target is at the top of the new list. 

A change of target will be forced prematurely if one of the following occurs: 
\begin{itemize}
\item  A `CHANGE' command is received from the supervisor. This would typically happen if an operator wanted to force an immediate change of target, usually following a change to the desired target separation.
\item Several  detector  pixels  saturate  repeatedly  (with  enough  tolerance  to  allow brief peaks, e.g. due to cosmic rays, short flashes of torch light). This might be caused by an LGS collision or some other unexpected light source entering the field of view. 
\item Several data sets in a row are rejected due to poor centroid signal to noise ratio, which would typically happen if there was thin cloud in front of the target star.
\item The software is unable to locate the spot pattern in the images. The likeliest reason for this would be thick cloud.
\end{itemize}


\subsection{Data processing}
\label{sec:dataProcessing}

\begin{figure*}   
	\includegraphics[width=\columnwidth]{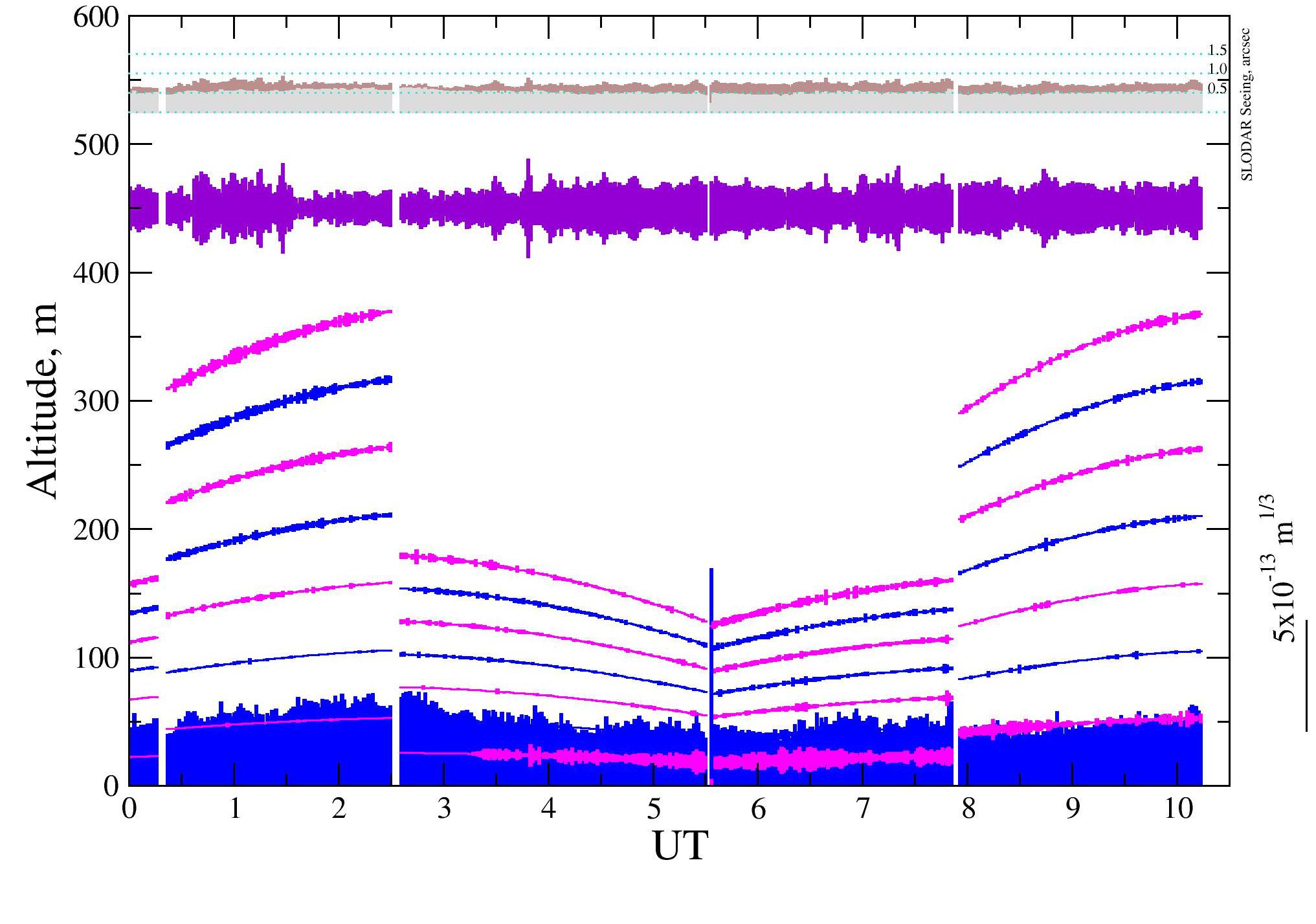}
	\includegraphics[width=\columnwidth]{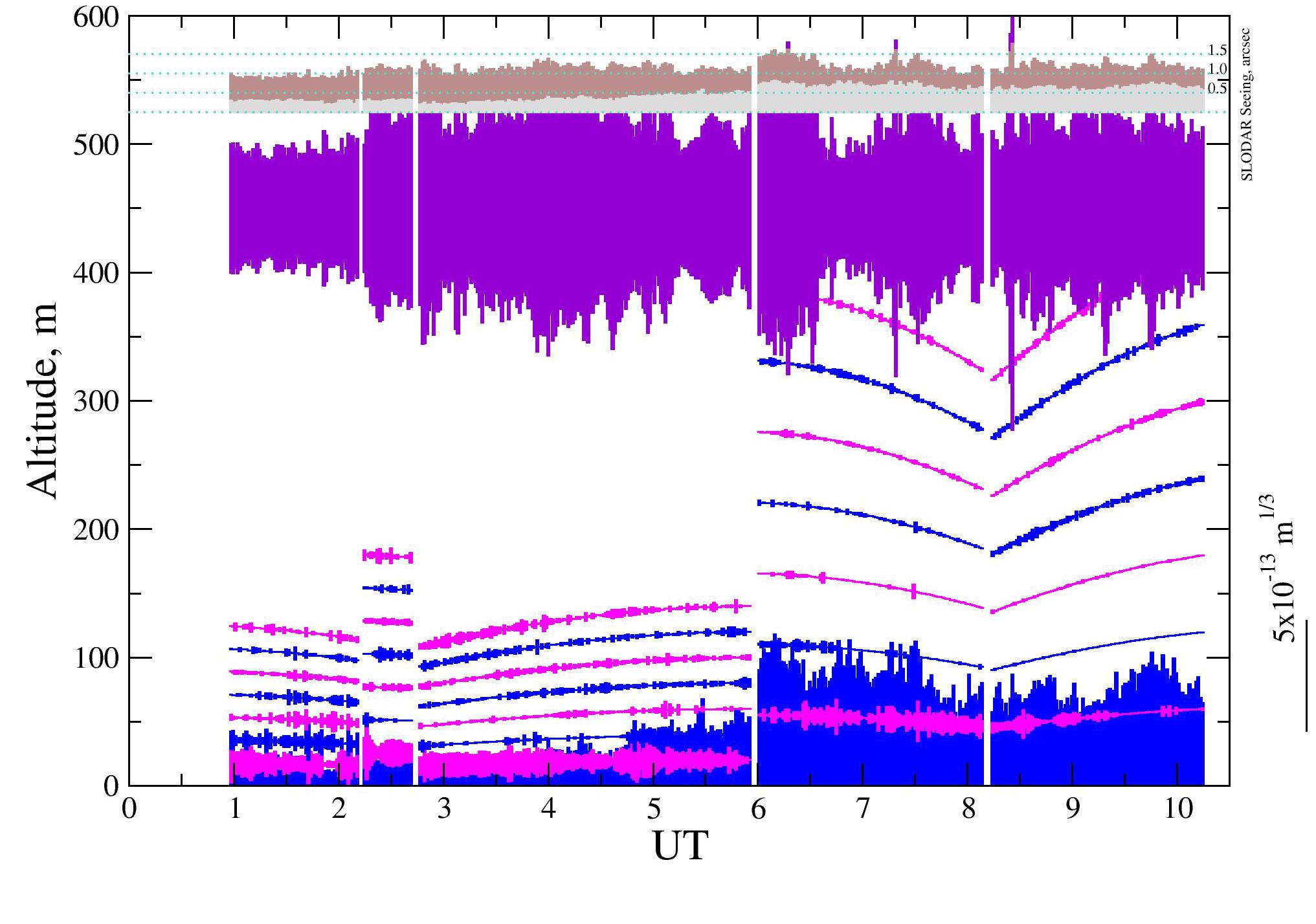}
	\includegraphics[width=\columnwidth]{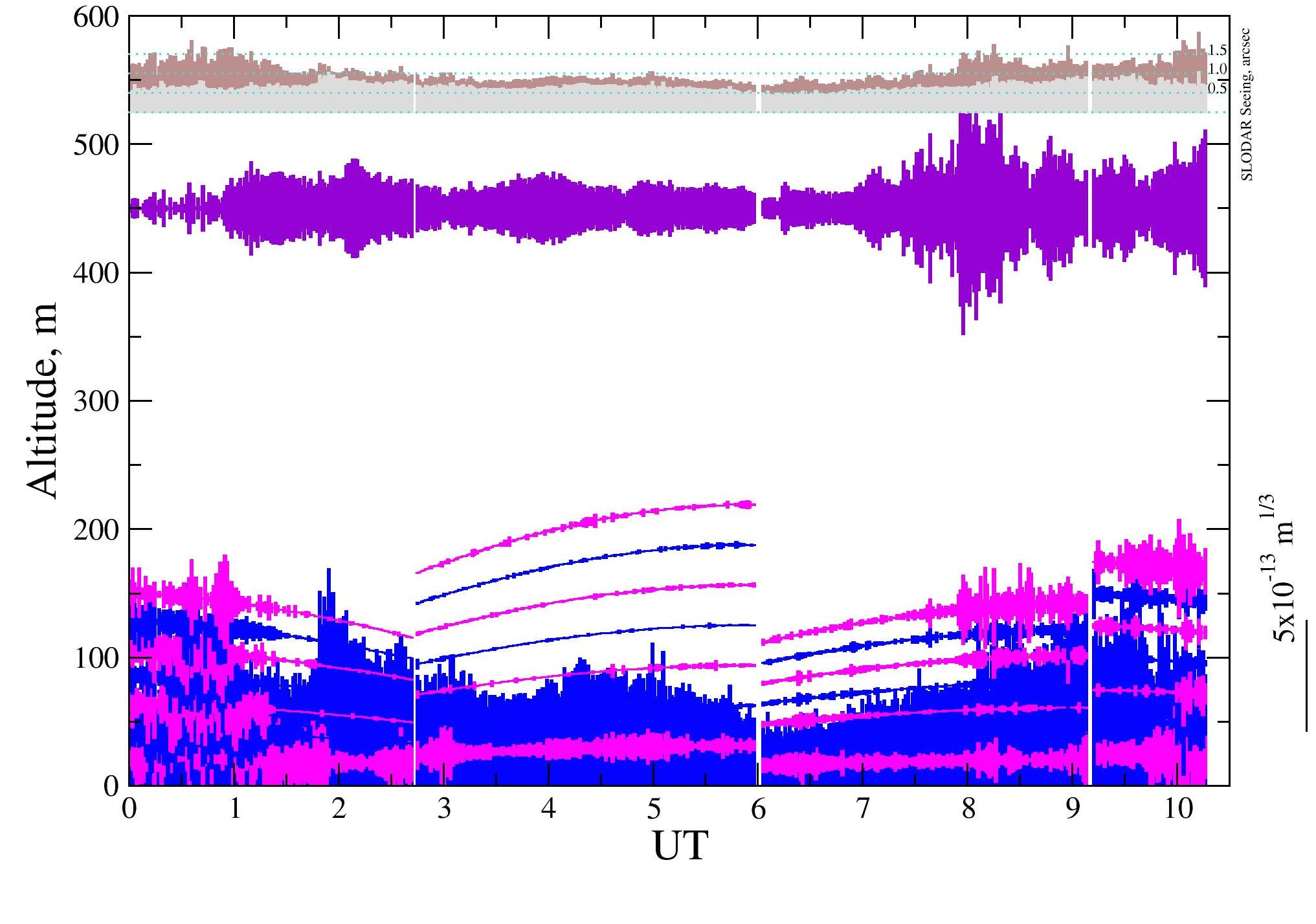}
	\includegraphics[width=\columnwidth]{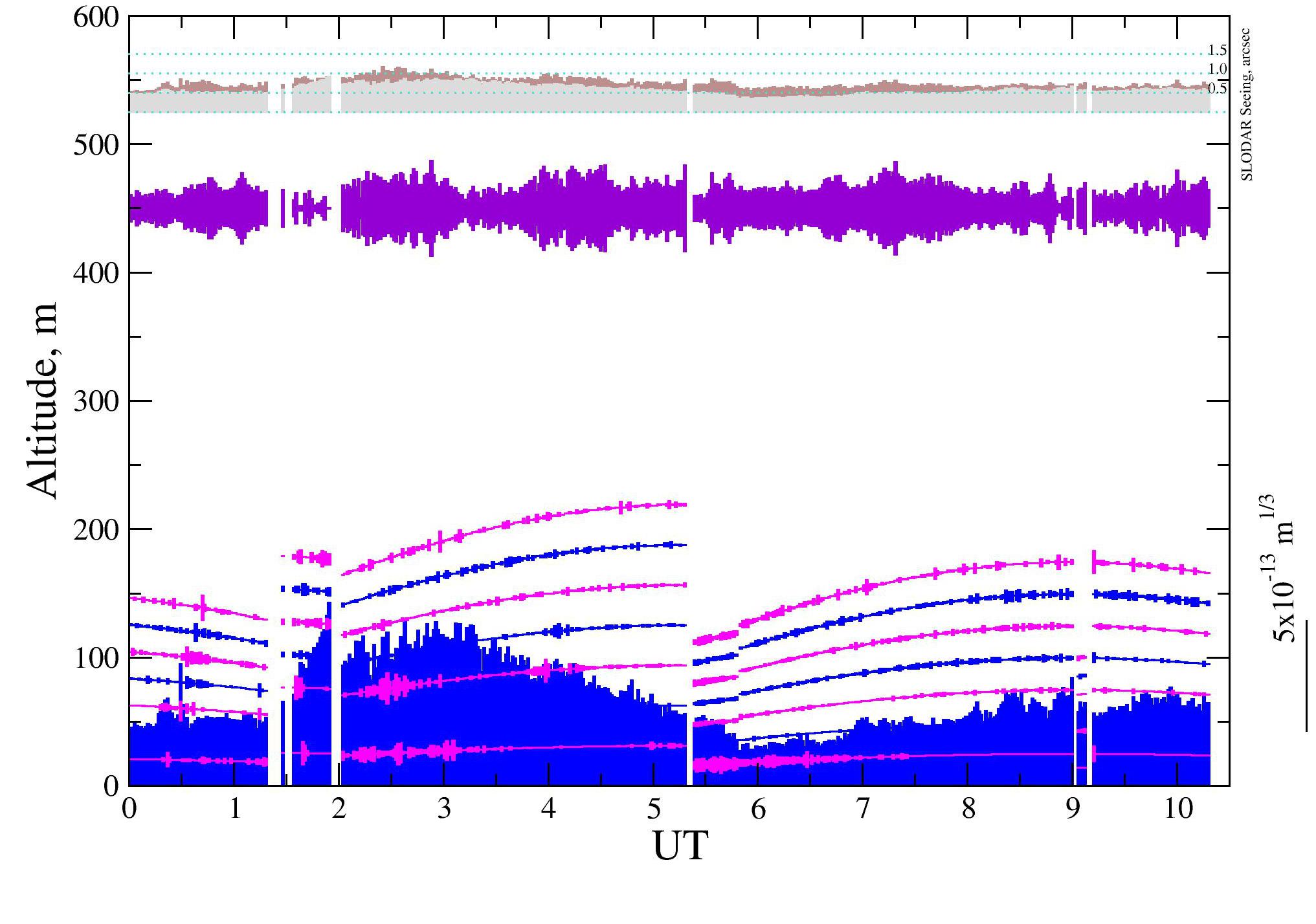}
	\caption{Example turbulence profile data (nights starting \hl{2015 April 15, 2015 April 17, 2015 April 22 and 2015 May 4}). The pink and blue traces represent the 8 resolution elements with alternating colour for clarity; each trace is centred at the height of the fitted layer and the thickness indicates the integrated $C_n^2 dh$ in the layer. Note that the traces change in height depending on the target separation \hl{(see figure~\ref{fig:geom})} and zenith angle. The purple trace shows the total integrated $C_n^2 dh$ above the maximum sensing height. The grey and brown traces show the seeing due to the ground layer and full atmosphere respectively.}
	\label{fig:exampleNightlyPlot}
\end{figure*}

WFS images are acquired and processed in `packets' of 1000 frames. Several packets are required to obtain sufficiently well-averaged slope covariances to recover the turbulence profile. There are two reasons for breaking up the dataset in this manner -- firstly to limit the amount of computer memory required to hold the images at any one time and secondly to limit the time between autoguiding updates.
An  image  packet  consists  of  two  sequences  of  WFS  images,  one  from  each EMCCD camera.  
First, pre-processing and quality control is carried out: 
\begin{enumerate}
\item Generate an average image from each WFS. 
\item Subtract the mean background value (measured at the corners of the frame) from each average image.
\item Attempt  to  locate  the  Shack-Hartmann spot  pattern  on  each  average  image.  If  this  fails  on either  image, for example due to clouds or incorrect pointing, reject  the  packet  (without  producing  any  autoguiding information). 
\item Measure  the position,  spacing  and  flux  of  each  average spot  pattern.  The   position is converted from detector coordinates to an offset in right ascension and declination; 
this is used  to  autoguide  the  telescope.
\item Quality control: if the position,  spacing, flux and overall rotation are not within the tolerances set for `good' data, reject the packet. 
\end{enumerate}

If the packet passes the quality control check, wavefront slopes are calculated from the image sequences as follows: 
\begin{enumerate}
\item Subtract the mean background level from every image in the packet. 
\item Select a region (the `centroiding box') around each spot in the pattern.  
\item For  each  frame  and  each  centroiding  box,  subtract  the  threshold  value (defined as 1/5 of the  mean  peak  value for that  spot)  from  the  sub-image. Round any negative values up to zero.  
\item Measure the centre of gravity of every sub-image. 
\item Referencing: For each spot, measure the mean centroid (averaged over all frames in the packet). Subtract this  position from each individual centroid so that the centroid is zero-mean, since the atmospheric wavefront slope information is contained in the deviation of the spot positions from their mean value.
\item Tip/tilt  subtraction:  For  each  frame,  measure  the  mean  x  and  mean  y centroid  over  all  of  the  spots.  Subtract these
so that common motion is removed from the centroid sequences. The purpose of this is to remove telescope wind-shake and tracking errors.
\item Calculate  the  spatial  auto-  and  cross-covariances  of  the  centroids as described by \citet{Butterley06b}. 
\item Append  the  auto-  and  cross-covariance arrays (three  arrays  in total: one auto-covariance array for each WFS and the cross-covariance  between the two WFSs) to a queue, until the queue contains enough packets to retrieve the turbuulence profile.
\end{enumerate}


Once a series of 6  centroid  packets 
has been accumulated, profile fitting proceeds as follows: 
\begin{enumerate}
\item Covariance preparation: Average the auto- and cross-covariance arrays in the queue to  obtain two auto-covariance maps, one for each WFS, and a single cross-covariance map.  Multiply each covariance map by the  image  scale  squared  and  divide  by  the  airmass so that the covariances are in units  of arcsec$^2$ at zenith. Fitting a model that is also in units  of arcsec$^2$  will then yield correctly-scaled zenith $C_n^2 dh$ values.
\item Estimate the integrated  seeing  ($r_0$):  Fit  a  Kolmogorov  model  to  the  auto-covariance  map  with  the  noisy  central  (variance)  peak  value excluded.  Do  this  separately  for  the  two  WFSs,  each  yielding  an  $r_0$  estimate.  
\item Estimate  noise  and  temporal  error:  Fit  a  non-Kolmogorov  model  to  each  auto-covariance  map  with  the  noisy  central  (variance)  value  excluded,  varying the exponent in the power spectrum, $\beta$, to obtain the value that gives the best fit.  The  centroid  noise  is  the  difference  between  the  measured  variance  and  that  predicted  by  the  best-fit  model.  Significant deviation from the Kolmogorov value (11/3)  indicates that the  dataset  is  poorly-averaged or that there is strong local non-Kolmogorov turbulence. We choose a threshold value of 3.4 -- if the power low exponent is smaller than this the dataset is discarded. See section~\ref{sec:beta} for a more detailed discussion.
\item Fit  a  set  of  Kolmogorov  response  functions  to  the  cross-covariance using a non-negative least squares algorithm. These  yield  $C_n^2 dh$  at  a  series  of  altitudes  corresponding to integer spatial offsets in the covariance map. 
\item Unresolved  $C_n^2 dh$:  Subtract  the  directly-sensed  integrated  $C_n^2 dh$  (from  the  profile fit) from the total integrated $C_n^2 dh$ (from the integrated seeing fit) to  estimate  the  integrated  $C_n^2$  above  the  maximum  sensing  altitude  of  the  instrument. 
\end{enumerate}

An  ASCII  format  file,  with  one  data  row  per  profile  measurement,  forms  the  main  data output from the SL-SLODAR. The main outputs are listed in Table~\ref{tab:slodarOutput}. Additional  data  recorded  to  archive  include  the  raw  centroid  data  for  both  WFSs  and  resulting  cross-covariance  values and other  diagnostic  data. Raw images are not saved as the volume of data would be too large.

\begin{table}
	\centering
	\caption{\label{tab:slodarOutput}Main outputs from a single SL-SLODAR measurement.}
	\begin{tabular}{lr} 
		\hline
		UT \\
		Target name \\
		Elevation \\
		Azimuth \\
		Airmass \\
		Flux ($\times 2$) \\
		Flux variance ($\times 2$) \\
		Centroid noise fraction ($\times 2$) \\
		Fried parameter, $r_0$ \\
		Kolmogorov criterion, $\beta$ \\
		Bin depth, $\delta h$ \\
		Turbulence strength in each layer, $C_n^2 dh$ ($\times 8$) \\
		Unresolved turbulence strength $C_n^2 dh$ \\
		\hline
	\end{tabular}
\end{table}

\subsection{Post-processing: surface layer model}
\label{sec:pipeline}

The profile of the turbulence in the first 500m varies greatly, as can be seen from the 
examples in figure~\ref{fig:exampleNightlyPlot}. However, we note that in nearly all cases there is a substantial  
surface layer contribution. This is seen as a strong signal in the first SL-SLODAR resolution 
element, centred at the telescope level. In many cases the second bin of the profile 
is relatively weak, suggesting that the scale height of the surface layer turbulence is only a  
few metres. Typically, this surface layer contribution is only clearly resolved in SL-SLODAR 
data with the highest vertical resolution (around 10~m), i.e. for observations of target stars with the 
largest separations (around 12~arcmin). 



This section describes an extension to the data processing pipeline, summarised by \citet{ButterleyAO4ELT4}, to provide an estimate of the turbulence above the height of the UT domes even when the surface layer is not resolved.
\begin{enumerate}
\item To find an appropriate model for the surface layer turbulence, the data from the prototype SL-SLODAR were used. That instrument operated with wide target angular separations and hence gave higher vertical resolution of the surface layer. The data with the largest target separations and for relatively low target elevations were selected, in order to resolve the surface layer turbulence as much as possible.
\item The prototype SL-SLODAR data were then fitted using an exponential model of the form
\begin{equation}
	C_n^2 (h) = A  \exp \left( \frac{- h}{h_0} \right) ,
\end{equation}
where $h$ is the height above the ground and $h_0$ and $A$ are constants. A combination of two such exponential components has previously been used to model the turbulence profile at Cerro Pach\'on \citep{Tokovinin06}.
Values of $A$ and $h_0$ were fitted to each profile in turn and, from the distribution of $h_0$ values, the optimum scale height for the model was found to be $h_0 = 5$~m.
\item The facility SLODAR data (2014 -- present) were then re-cast onto a regular vertical grid. The method is described in detail in appendix~\ref{sec:appendix} and is summarised as follows:   
	\begin{enumerate}
	\item Start with the $C_n^2$ profile obtained as described in section~\ref{sec:dataProcessing} (8 sensed layers, variable altitude depending on target separation and zenith angle). 
	\item An exponential surface-layer component was calculated using the model defined from steps (i) and (ii). This was re-binned onto the actual vertical resolution of the SL-SLODAR observation and scaled in strength according to the $C_n^2 \dif h$ value of the first SL-SLODAR bin. In the event that the target separation was wide enough for the surface layer model to extend into the second bin and exceeded the observed $C_n^2 \dif h$ in that bin the surface layer model strength was reduced until this was rectified.
	\item The surface layer component, as calculated in (b) was subtracted from the original SL-SLODAR profile. The remainder of the SL-SLODAR profile was re-binned onto a 1~m vertical profile using the known (triangular) SL-SLODAR response/weighting functions centred at the altitude of each original SL-SLODAR vertical bin.
	\item The final $C_n^2$ profile, on a regular 1~m vertical grid, was the sum of the surface layer component from (b) plus the result of (c). 
        \end{enumerate}
\end{enumerate}

\todo{Refer to appendix?}


\section{Convergence and low wind speed behaviour}
\label{sec:beta}

\todo{RWW's points?}

In this section we discuss how to diagnose and interpret poorly-converged slope covariance measurements.

As noted in section~\ref{sec:dataProcessing}, a generalized power spectrum is fitted to the measured slope autocovariance as a test of the data quality. We adopt the generalized phase power spectrum expression described by \cite{Nicholls95},
\begin{equation}
I_{\phi}(\kappa) = \frac{A_{\beta} \kappa^{-\beta}}{\rho_0^{\beta-2}} \;\;\;\;\;\;\;\; (2 < \beta < 4)
\end{equation}
where $\rho_0$ is analogous to $r_0$ and $A_\beta$ is a constant chosen such that the piston-subtracted wavefront variance over a pupil diameter $D=\rho_0$ is equal to 1~rad$^2$.

One expects well-averaged Kolmogorov turbulence to yield a power spectrum exponent of $\beta = -11/3$. One also expects this of well-averaged Von Karman turbulence for SLODAR on a 0.5~m telescope, since the outer scale is generally considerably larger than the aperture and global tip/tilt is excluded from the analysis. Tip and tilt are the modes that are most sensitive to the outer scale so, without them, we are in a regime where Von Karman turbulence is indistinguishable from Kolmogorov turbulence provided $L_0 >> D$.  \hl{(see e.g. \citet{Winker91})}

The power spectrum exponent, $\beta$, is measured by fitting theoretical autocovariance functions for a range of $\beta$-values to the measured autocovariance (for a single star).
Measuring a non-Kolmogorov power spectrum exponent, $\beta < -11/3$, can have two explanations:
\begin{enumerate}
\item The turbulence is Kolmogorov/Von Karman but the wind speed is too slow or packet length is too short for the slope covariances measurements to average fully.
\item The turbulence is not Kolmogorov/Von Karman. In general the free atmosphere is accepted as being Von Karman \todo{Ref(s)? Amokrane?} but this may not be true for local turbulence in and around the SL-SLODAR enclosure.
\end{enumerate}

\begin{figure}
	\centering
	\includegraphics[width=0.9\columnwidth]{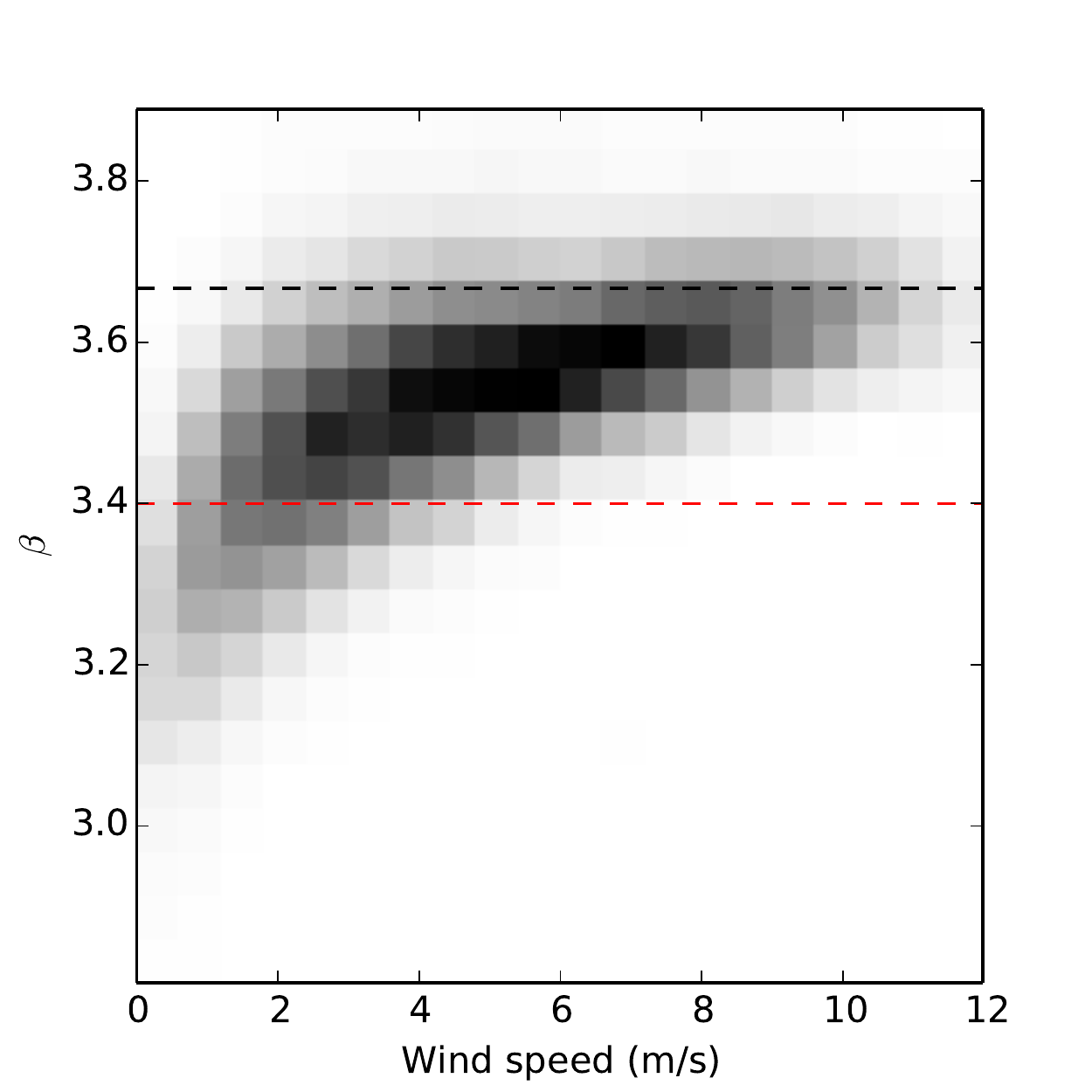}
	\caption{Density plot showing $\beta$ measured from SL-SLODAR as a function of wind speed 10~m above the ground from the meteo mast. The black broken line shows the Kolmogorov case and the red broken line shows the $\beta = 3.4$ threshold.}
\label{fig:windVsBetaData}
\end{figure}

In practise we frequently observe values of $\beta$ that are lower than $-11/3$. There is a clear dependence on wind speed, as seen in figure~\ref{fig:windVsBetaData}. It is common to observe $\beta < 3.4$ 
when the wind speed measured 10~m above the ground is less than $\sim 3$~m/s.

\subsection{Effect of increased packet size}
\label{sec:betaPacketSize}

\begin{figure}   
	\includegraphics[width=\columnwidth]{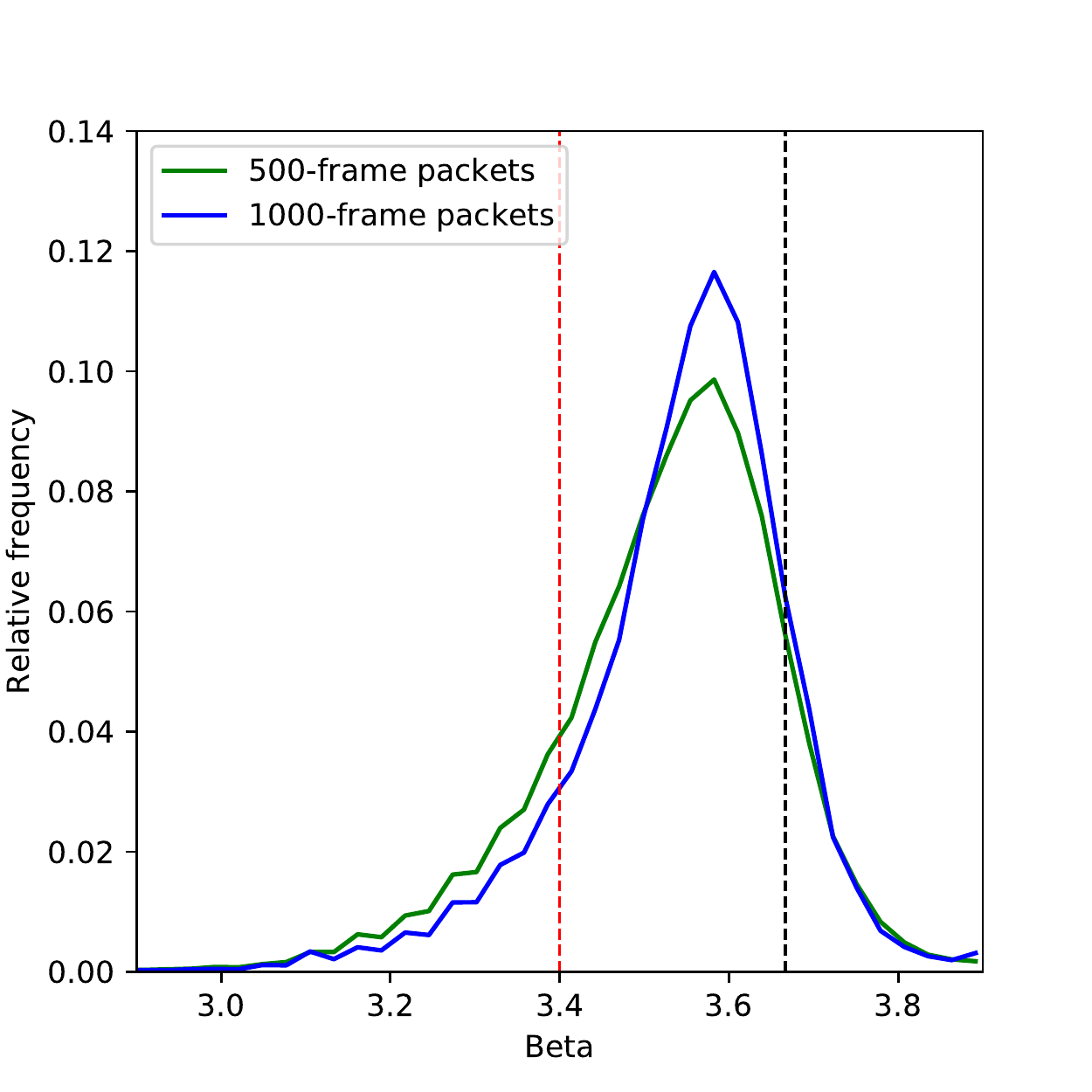}
	\caption{Distributions of $\beta$ values before and after increasing the packet length from 500 frames to 1000 frames. The black broken line shows the Kolmogorov case and the red broken line shows the $\beta = 3.4$ threshold.}
	\label{fig:betaVsPacketLength}
\end{figure}




The packet size was increased from 500 frames to 1000 frames on 2016-01-26.
Figure~\ref{fig:betaVsPacketLength} shows the distributions of $\beta$ values before and after this change. 
Doubling the packet length had the effect of increasing $\beta$ at low wind speeds but only by a small amount.
The fraction of data points for which $\beta > 3.4$ has increased from 83\% (for 500-frame packets) to 88\% (for 1000-frame packets).



\subsection{Temporal averaging simulation}
\label{sec:betaSim}

In this section we demonstrate via Monte Carlo simulation that the observed $\beta$-values can not be explained simply by insufficient averaging of freely moving turbulence outside the dome.
First we consider what behaviour we expect to see if we assume the instrument sees only `well-behaved' Von Karman turbulence.

Each SL-SLODAR profile is currently generated from 5 packets of data, each 1000 frames long (500 frames prior to \hl{2016 January 26}). The camera frame rate is 57.6~Hz 
so each packet has a duration of 17 seconds. 
Each packet is reduced separately, so the mean spot positions (i.e. `static' aberration) are calculated over the 17 seconds and subtracted. We expect to see artificially small 
$\beta$ if the turbulence does not change enough for the `static' aberration to average out to approximately zero in this time \citep{ButterleyATTA}.

The temporal averaging effect was modelled using a Monte Carlo simulation, assuming Taylor frozen flow and a 30~m outer scale (so essentially indistinguishable from Kolmogorov as seen by our 0.5~m aperture) to generate artificial packets of slopes of the correct duration for different wind speeds. These were reduced in exactly the same way as real data (averaging over several simulated packets) to yield autocovariances for a single ideal layer. 

\begin{figure}   
	\includegraphics[width=\columnwidth]{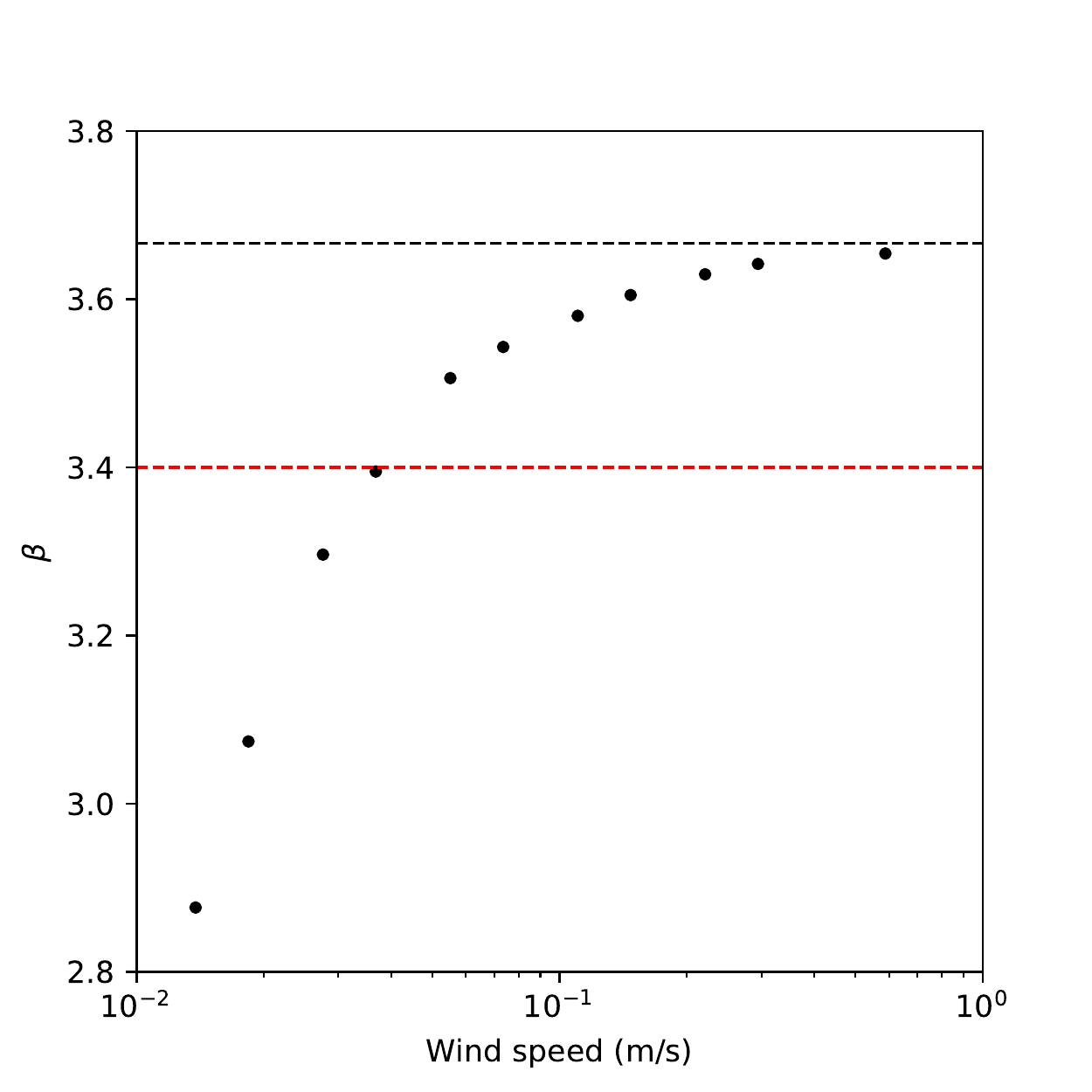}
	\caption{Effect of (Taylor frozen flow) wind speed on $\beta$ (where $\beta$ is the exponent in the turbulence power spectrum). The black broken line shows the Kolmogorov case and the red broken line shows the $\beta = 3.4$ threshold.}
\label{fig:windVsBetaSim}
\end{figure}

The value of $\beta$ was fitted to each simulated autocovariance. The results are shown figure~\ref{fig:windVsBetaSim}. As expected, $\beta$ is low for slow wind speed and consistent with Kolmogorov (black broken line) for high wind speed, but there is a major discrepancy between the $\beta$-wind speed relation here and that observed at Paranal. For simulated data $\beta$ drops below the threshold of 3.4 (red line) at a wind speed of 0.035~m/s in the simulation, compared to $\sim 3$~m/s at Paranal.

The following factors may be contributing to this discrepancy:
\begin{enumerate}
\item The Paranal wind speed is measured 10~m above the ground. The wind speed near the telescope (2~m above the ground) is probably slower most of the time. However, one certainly would not expect it to be slower by a factor of $\sim$100; a difference of more than 10\% seems unlikely.
\item The SL-SLODAR suffers from a significant dome seeing contribution due to local heat sources e.g. the EMCCD cameras (each of which has a maximum power draw of 16~W). One would expect this to have a pronounced effect when the wind speed is too slow to flush the warm air out of the dome.
\item \hl{The SL-SLODAR suffers from turbulence generated by the numerous heat sources in the ASM hut (see section~\ref{sec:instDesign}). If this were the case, one would expect $\beta$ to depend strongly on wind direction i.e. the turbulence should predominantly be non-Kolmogorov when the wind blows across the ASM hut towards the SL-SLODAR. In practice $\beta$ is seen to vary only weakly with wind direction so this is at most a secondary effect.}
\item The Kolmogorov frozen flow model for the surface layer (outside the dome) may be inadequate in the low wind speed regime. 
\end{enumerate}

Of these possibilities, the second seems likely to be the most significant effect. As noted in section~\ref{sec:instDesign}, the dome sides are louvred but they restrict air flow through the enclosure more than they would if they were completely open. This compromise was necessary to protect the instrument from the elements.



\subsection{Implications}

As mentioned in section~\ref{sec:dataProcessing}, the turbulence profile is fitted by assuming a Kolmogorov model for the turbulence at all altitudes. In the case where $\beta < 11/3$ at the ground, this model fits the data poorly and tends to lead to the turbulence strength being over- or underestimated in other resolution elements. If one did not enforce positivity in the profile fit the effect of fitting too broad a peak at the ground would be to produce unphysical negative $C_n^2$ values in the first resolution element above the ground.

In order to ensure the integrity of the SL-SLODAR profiles, data taken in the regime where $\beta$ is below the empirically-determined threshold of 3.4 is deemed to be unreliable and is flagged as invalid.











\section{Statistical results and discussion}
\label{sec:stats}

Paranal SL-SLODAR data from April 2016 onwards is publicly available from the ESO `Paranal Ambient Query Forms' web page.\footnote{\url{http://archive.eso.org/cms/eso-data/ambient-conditions/paranal-ambient-query-forms.html}}
Data from prior to April 2016 is available from the authors on request.

Throughout most of 2014 and 2015 the system predominately observed targets 
with the widest separations available. This permitted a statistical characterisation of the 
turbulence strength close to ground level. From December 2015 onwards the system has been configured to select 
targets with narrower separations (lower altitude resolution) in order to map the turbulence profile 
up to an altitude of approx. 500~m, matching the range of altitudes targeted for correction by the 
AOF system.

\subsection{Raw statistics}
\label{sec:RawStatistics}

Table~\ref{tab:observations} shows the numbers of nights that have been observed and the numbers of individual profile measurements accumulated in each month of the year.


\begin{table}
	\centering
	\caption{\label{tab:observations}Number of SL-SLODAR nights/observations by month (January 2014 -- September 2018).}
    \begin{tabular}{ccc}
    \hline
    Month & Nights & Individual \\
    & observed & observations \\
    \hline
    Jan & 73 & 13330 \\
    Feb & 69 & 9576 \\
    Mar & 83 & 12996 \\
    Apr & 103 & 19387 \\
    May & 78 & 13115 \\
    June & 66 & 6669 \\
    July & 90 & 15936 \\
    Aug & 46 & 8785 \\
    Sept & 43 & 7574 \\
    Oct & 42 & 10383 \\
    Nov & 96 & 17556 \\
    Dec & 86 & 15085 \\
    \hline
    TOTAL & 875 & 150392 \\
    \hline
    \end{tabular}
\end{table}

\begin{figure}   
 \includegraphics[width=\columnwidth]{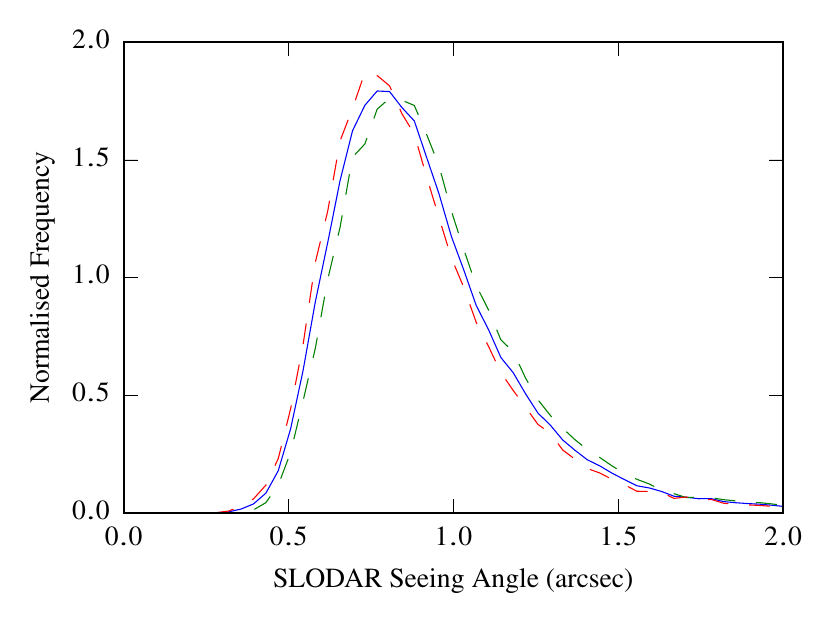}
 \caption{Normalised frequency distribution of measured seeing angle values for SL-SLODAR 
 -- see section~\ref{sec:RawStatistics}. The solid line indicates the whole SL-SLODAR database 
 (median value 0.861~arcsec). The broken lines show the seasonal variation, with a median value 
 of 0.837 for the summer months (October to March, red line) and 0.889 for the winter months 
 (April to September, green line). }
 \label{fig:FWHM_Hist}
\end{figure}

The frequency distribution of seeing angle values for the full SL-SLODAR data set is shown in figure~\ref{fig:FWHM_Hist}, comprising a total of 155696 individual measurements over 932 nights 
between \hl{2013 Sep 21 and 2018 Sep 19}. We find a median value for the seeing angle of 0.861~arcsec. 
This value is significantly larger than the median seeing estimated from the DIMM seeing monitor 
at Paranal: \hl{we find a median seeing angle of 0.743~arcsec for the Paranal DIMM data used in this study (see section~\ref{sec:DIMM})}. We attribute this difference to effect of the surface layer of turbulence 
and the relative height of the SL-SLODAR and DIMM monitors, as discussed in the following section. 
We find a significant seasonal variation in the SL-SLODAR seeing measures, with a median value of 
0.837~arcsec for the summer months (October to March) and 0.889~arcsec for the winter months 
(April to September). We show in section~\ref{sec:ExponentialModel} that this variation is 
associated with the surface layer turbulence, with no significant seasonal variation in the 
integrated turbulence strength above 50~m. 

\subsection{Exponential surface layer model}
\label{sec:ExponentialModel}
\begin{figure}    
 \includegraphics[width=\columnwidth]{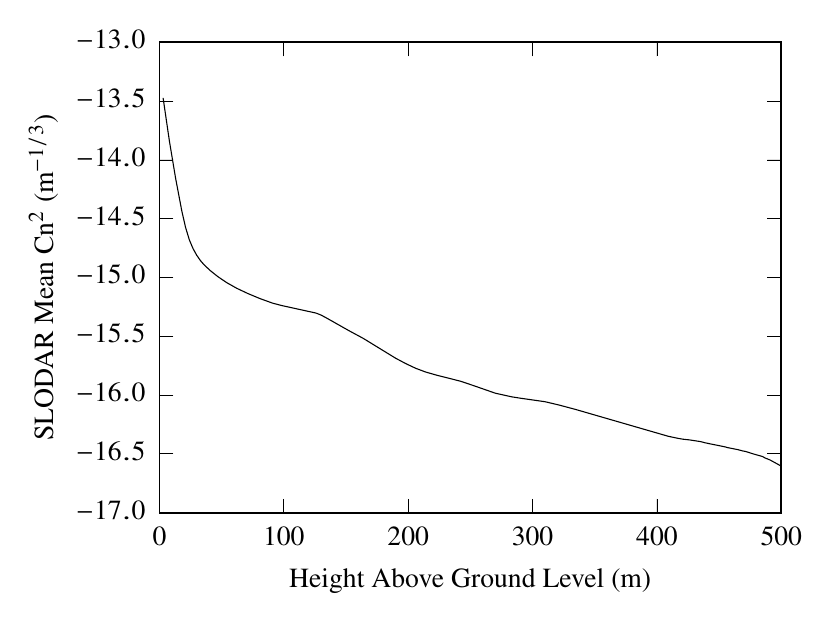}
 \caption{Mean optical turbulence profile measured by SL-SLODAR, from a total of 155696 individual 
 profile measurements over 932 nights between 2013 and 2018. The data have been processed using 
 the analysis described in section~\ref{sec:pipeline} which includes the exponential surface layer 
 model component}
 \label{fig:MeanCn2}
\end{figure}

\begin{figure}    
 \includegraphics[width=\columnwidth]{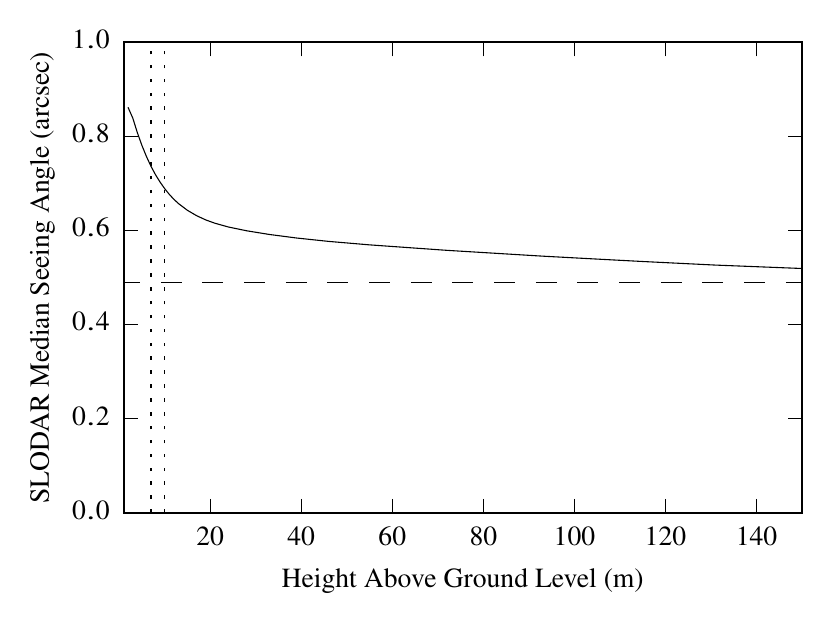}
 \caption{Median seeing angle versus altitude, based on the entire data set of SL-SLODAR profiles, 
 processed using  the analysis described in section~\ref{sec:pipeline} and including the 
 exponential surface layer model component. The horizontal broken line indicates the median seeing 
 angle for altitude 500~m (0.481~arcsec). 
 The vertical dotted lines indicate the altitudes 
 of the DIMM seeing monitor (7~m) and of the observing floor of VLT UT4 (10~m) above ground level.}
 \label{fig:FWHM_Median_H}
\end{figure}

Figure~\ref{fig:MeanCn2} is the mean optical turbulence profile for the whole SL-SLODAR data set. 
This is calculated for the data processed using the analysis described in section~\ref{sec:pipeline} 
and includes the exponential surface layer component, which dominates the profile in the first 
30~m above the ground. Figure~\ref{fig:FWHM_Median_H} plots the median seeing angle value measured  
by the SL-SLODAR as a function of altitude above ground level, from the height of the SL-SLODAR 
monitor at 2~m.  Assuming infinite outer scale, the seeing angle is given by
\begin{equation}
	\theta = 0.98 \ \lambda/r_{0},
	\label{eq:seeingangle}
\end{equation}
where $r_{0}$ is the value of the Fried parameter corresponding to the integrated turbulence strength above the observing altitude, and $\lambda$ is the observing wavelength, assumed to be 500~nm. \citep{Sarazin90}.

The fraction of the turbulence strength associated with the exponential surface 
layer component is usually substantial, so that we see a large and rapid decrease in the median seeing 
with increasing altitude, over the first 20~m. \hl{This is consistent with the findings of \citet{Lombardi10}.}

\begin{figure}
 \includegraphics[width=\columnwidth]{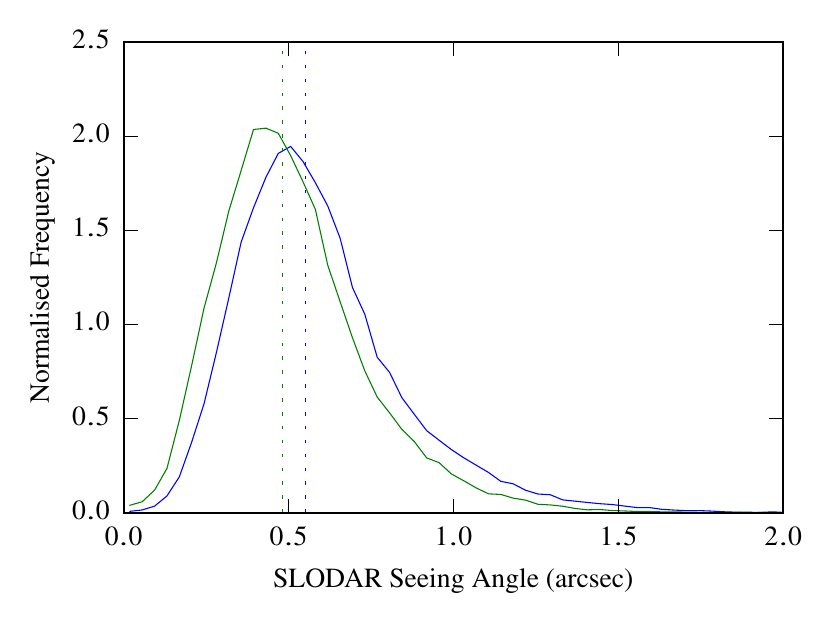}
 \includegraphics[width=\columnwidth]{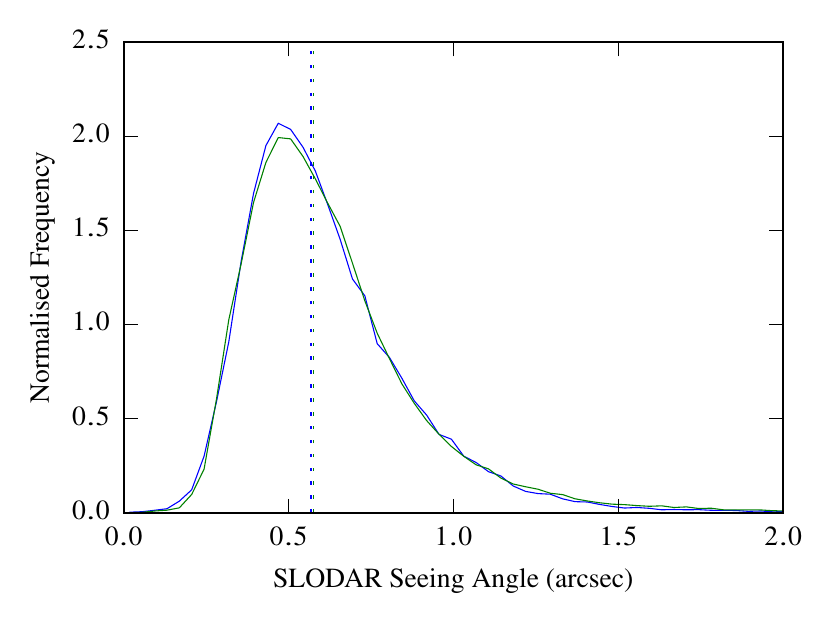}
 \caption{Normalised frequency distributions of SL-SLODAR seeing angle values. Each panel shows summer (green) and winter (blue) months. Upper: turbulence below 50~m (summer median 0.481 arcsec, winter median 0.552 arcsec);  lower:  turbulence above 50~m (summer median 0.568 arcsec, winter median 0.575 arcsec). The median value of each distribution is shown by a vertical dotted line in the same colour.}
 \label{fig:FWHM_Hist_BelowAndAbove50m}
\end{figure}

We find a significant seasonal variation in the strength of the surface layer turbulence. 
Figure~\ref{fig:FWHM_Hist_BelowAndAbove50m} (upper) shows the frequency distribution of the seeing angle  
associated with the surface layer of turbulence (only), up to altitude 50~m, for summer (October to March) 
and winter (April to September) months. The median seeing corresponding to the surface layer 
turbulence is 0.481~arcsec in the summer months 0.552~arcsec in the winter. Above the surface layer 
(altitude above 50m) we find no significant seasonal variation in the integrated turbulence strength, 
with a median seeing value of 0.568~arcsec for the summer months and 0.575~arcsec 
for the winter months. Figure~\ref{fig:FWHM_Hist_BelowAndAbove50m} (lower) shows the frequency distributions 
of the seeing angle for the integrated turbulence above 50~m, for the summer and winter months.

The effect of this strong, thin, surface layer turbulence must be taken into account when 
estimating the seeing relevant to the UT and other telescopes at the Paranal site. 
The height above ground level of the observing floor of the UT is 10~m. 
The effective height of the UT for calculating the fraction of the surface layer that will 
contribute to the seeing is not certain, since the exact effects of the UT enclosure 
on the surface layer of turbulence local to the telescope are not known. However for this 
estimation we assume that the exponential profile of the surface layer used for the SL-SLODAR 
analysis is appropriate, and that turbulence below the height of the observing floor does not 
contribute to the seeing of the UT. For altitude = 10~m, we find a median seeing value of 
0.689~arcsec from the exponential model fit to the full SL-SLODAR data set.

\subsection{Comparison with DIMM seeing monitor}
\label{sec:DIMM}

In order to explore whether the exponential model fit to the lowest-altitude turbulence strength 
in the SL-SLODAR data provides an accurate estimate of the seeing as a function of altitude, we can 
compare to contemporaneous measurements from the DIMM seeing monitor of the ASM at Paranal. 
DIMM measures the total integrated optical turbulence strength over all altitudes, via the differential 
image motion method \citep{Sarazin90}. 
The DIMM is located on a tower at a height of  7~m, on the eastern edge of the VLT observing platform, 
approximately 80~m south of the location of the SL-SLODAR. 

\begin{figure}    
 \includegraphics[width=\columnwidth]{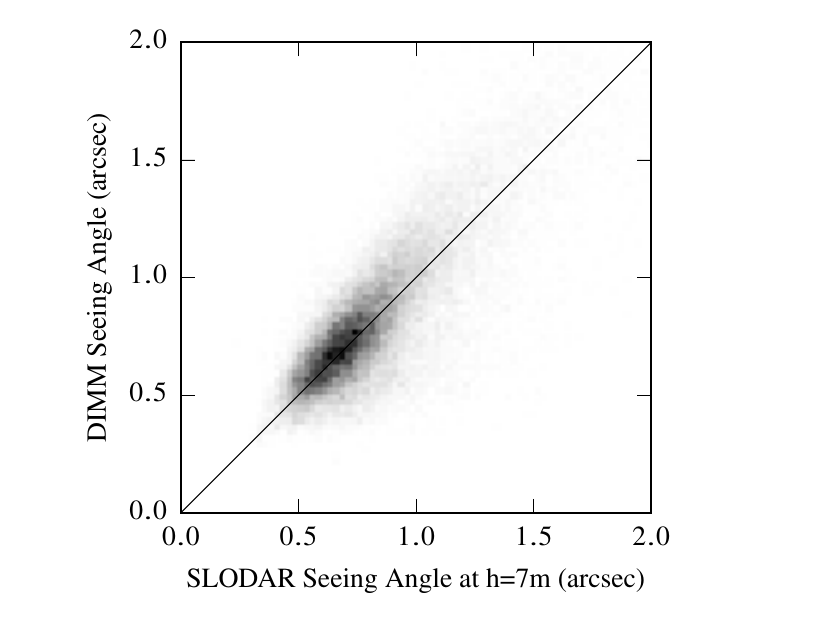}
 \caption{Comparison of SL-SLODAR seeing angle values for altitude 7~m 
 and
 contemporaneous DIMM seeing measurements (same data as included in the frequency plots, figure~\ref{fig:FWHM_Hist_DIMM}). Correlation coefficient $= 0.808$. The black line shows the $y=x$ case for reference.}
 \label{fig:SLOD_DIMM}
\end{figure}
\begin{figure}
 \includegraphics[width=\columnwidth]{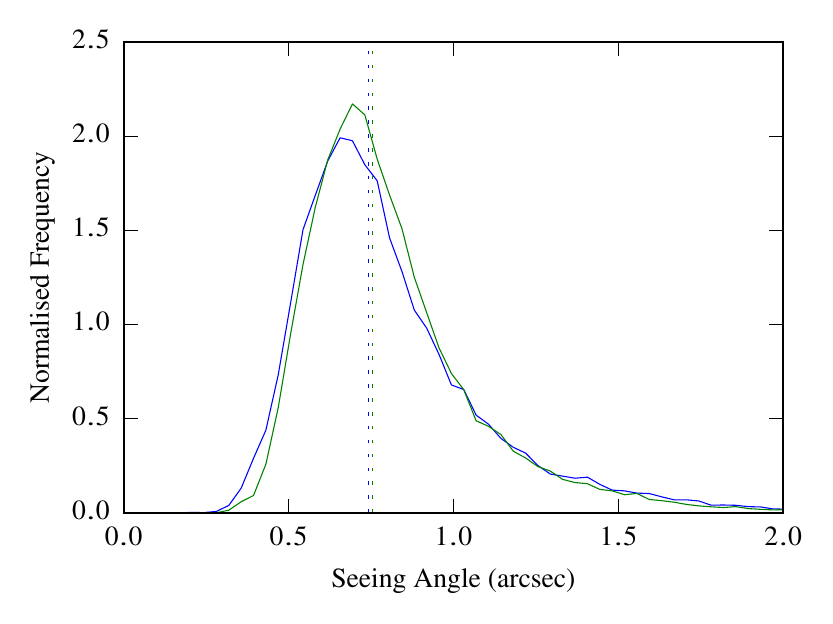}
 \caption{Normalised frequency distribution of SL-SLODAR seeing angle values for the height of the Paranal DIMM seeing monitor (7~m) (green line, median 0.755 arcsec) and for contemporaneous DIMM seeing measurements (blue line, median 0.743 arcsec),  total of 33722 contemporaneous measurements on 353 nights between 2016 and 2018. The median value of each distribution is shown by a vertical dotted line in the same colour.}
 \label{fig:FWHM_Hist_DIMM}
\end{figure}

Figure~\ref{fig:SLOD_DIMM} shows a comparison of the seeing angle values measured by the DIMM 
and by the SL-SLODAR, assuming the exponential model and at the height of the DIMM, for 
contemporaneous data from the two instruments. These comprise a total of 33722 contemporaneous measurements 
made on 352 nights between 2016 April 5 and 2018 September 19. Figure \ref{fig:FWHM_Hist_DIMM} 
shows the corresponding frequency distributions of the seeing values for SL-SLODAR and DIMM, 
for the contemporaneous data. We compare each SL-SLODAR measurement with the mean of all DIMM values 
recorded within 3 minutes of the same time. We find a median value of the seeing angle of 0.755~arcsec for the 
SL-SLODAR at the height of the DIMM, and 0.743~arcsec for the DIMM itself, for the contemporaneous data. 
The correlation coefficient between the two data sets is 0.808. Given that the two seeing monitors are not co-located, they do not 
observe the same target stars, and that the observations were not perfectly synchronised in time, 
substantial scatter in the comparison can be expected. However, given the similarity of the distributions 
and median values, and the high degree of correlation found, we conclude that the exponential model fit 
to the SL-SLODAR data  provides an accurate estimate of the seeing at the altitude of the DIMM.

\subsection{Comparison with the image width of the VLT active optics wavefront sensor}

\begin{figure}
 \includegraphics[width=\columnwidth]{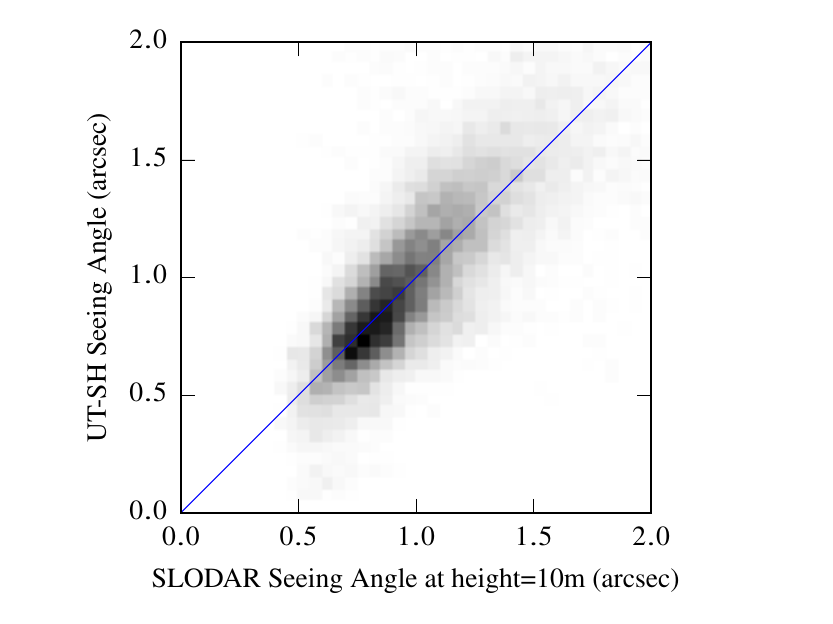}
 \caption{Comparison of SL-SLODAR seeing angle values for altitude 10~m and
 for contemporaneous seeing angle estimates from the Shack-Hartmann wavefront sensor of active optics system of UT1. Correlation coefficient $= 0.475$. The blue line shows the $y=x$ case for reference.}
 \label{fig:SLOD_UTSH}
\end{figure}

\begin{figure}   
 \includegraphics[width=\columnwidth]{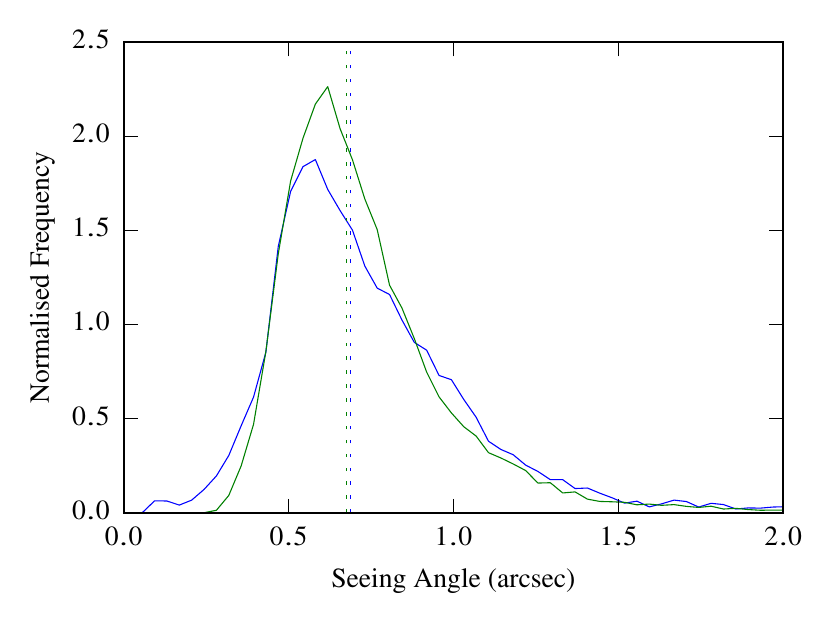}
 \caption{Normalised frequency distribution of SL-SLODAR seeing angle values for altitude 10~m (green, median 0.676 arcsec) and for contemporaneous seeing estimates from the  Shack-Hartmann wavefront sensor of the active optics system of VLT UT1 (blue, median 0.687 arcsec). The median value of each distribution is shown by a vertical dotted line in the same colour.}
 \label{fig:FWHM_Hist_UTSH}
\end{figure}

Figure~\ref{fig:SLOD_UTSH} shows a comparison of the seeing angle measured by the SL-SLODAR, assuming 
the exponential model and at the height of the UT primary mirror, with estimates of the seeing angle 
extracted from the Shack-Hartmann WFS of the active optical system of UT1, which we refer to as UTSH. 
The comparison includes a total of 28393 contemporaneous measurements from the two instruments 
on 297 nights between 2014 January 1 and 2015 December 31. We compare each SL-SLODAR measurement with 
the mean of all UTSH values recorded within 3 minutes of the same time. 
Figure \ref{fig:FWHM_Hist_UTSH} shows the frequency distributions of the UTSH and SL-SLODAR 
(corrected to height \ = \ 10~m) seeing angle values, for the contemporaneous data.

The active optics Shack--Hartmann comprises an array of 24 by 24 sub-apertures projected 
across the diameter of the telescope pupil, each with a projected width of 34~cm.
The VLT control system software produces a measurement of the median FWHM of the spots in the 
Shack-Hartmann pattern, for each wavefront sensor exposure of duration 30~seconds \citep{Martinez12}.  

The WFS spots of the UTSH have a diffraction--limited FWHM of 0.45~arcsec at the effective 
wavelength of the wavefront sensor (750~nm), which is convolved with the broadening of the spots 
due to the seeing. We therefore subtract 0.45~arcsec in quadrature from the reported FWHM values 
in order to estimate the seeing angle. 

The FWHM measurements from the UTSH are also affected by the finite spatial sampling of the 
Shack--Hartmann image by the pixels of the wavefront sensor detector (0.31~arcsec/pixel). 
However, as we do not have access to the details of the algorithm used, we are not able 
to model the effects of sampling on the output FWHM in detail. We estimate the size of 
the required correction as the fractional increase in the FWHM of a Gaussian function 
(representing the PSF of a wavefront sensor spot) when convolved with a square pixel 
response. 

Finally we scaled the FWHM values to their expected value at wavelength 500~nm, for comparison 
with SL-SLODAR, with the standard assumption that the seeing--limited FWHM scales as $\lambda^{1/5}$.

From the analysis of active optics image FWHM data we find a median seeing value of 0.687~arcsec, 
which is close to the median value of 0.676~arcsec for the contemporaneous SL-SLODAR data corrected 
to altitude = 10~m. The scatter in the comparison of seeing values is larger than for 
the comparison of SL-SLODAR with DIMM, with a correlation coefficient of only 0.475. 
This increased scatter may result in part from the larger physical separation (approximately 180~m) 
between SL-SLODAR and UT1, which are located on opposite sides of the Paranal observing platform. 
Furthermore the UTSH seeing estimate is likely to be slightly increased by any guiding errors 
or wind--shake of the telescope. On the other hand there will be a small reduction of the FWHM 
of the UTSH spots due to the effects of the outer scale of turbulence. These effects will all  
vary with time and will account for some of the scatter in the comparison with the SL-SLODAR seeing. 
However, we conclude that there is no large bias in the estimate of the UTSH seeing 
found from the SL-SLODAR data and therefore that we can usefully extend the SL-SLODAR model to estimate 
the performance of optimal GLAO correction for the UTs. 

\subsection{GLAO performance and the free--atmosphere seeing strength}
\label{sec:GLAOperformance}

The SL-SLODAR data can be used to estimate the best possible performance of GLAO correction for the UTs, 
in the hypothetical case where perfect AO correction can be applied to all aberrations due to 
optical turbulence up to a given height above the telescope - this is equivalent to the 
the seeing value at the corresponding height, found from figure~\ref{fig:FWHM_Median_H}. 

\begin{figure}
 \includegraphics[width=\columnwidth]{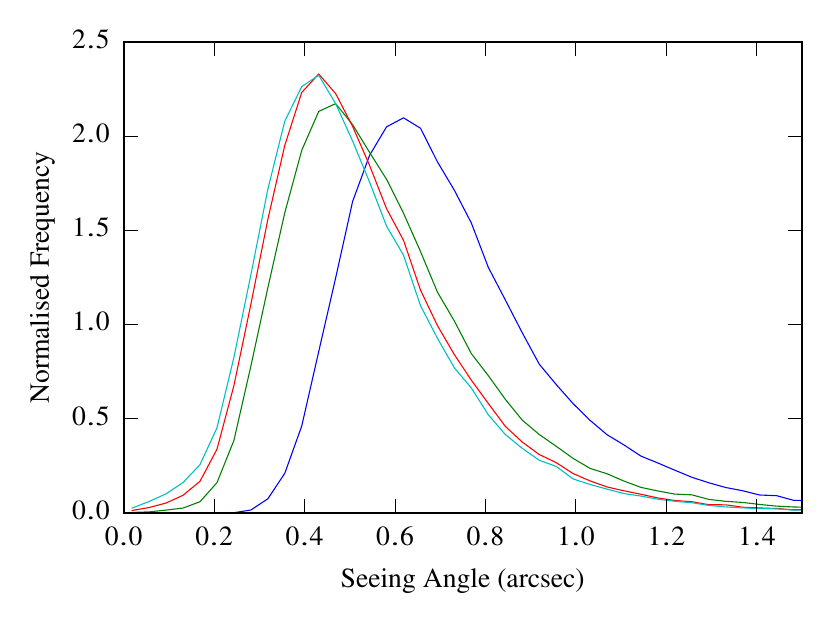}
 \caption{Normalised frequency distributions of SL-SLODAR seeing angle values for altitude 10~m (blue), 
 100~m (green), 250~m (red) and 500~m (light blue). Median values are 0.689 arcsec, 0.541 arcsec, 0.498 arcsec and 0.481 arcsec respectively.}
 \label{fig:FWHM_Hist_GLAO}
\end{figure}

Normalised frequency distributions are shown in figure~\ref{fig:FWHM_Hist_GLAO} for the 
median SL-SLODAR seeing angle value at the altitude of the observing floor (10~m) and at altitudes 
of 100~m, 250~m and 500~m.  The median seeing values (at wavelength 500~nm) from the SL-SLODAR data 
for these altitudes are 0.689~arcsec, 0.541~arcsec, 0.498~arcsec and 0.481~arcsec respectively.  
\todo{Add comment}

The relative contributions  of the ground layer and free--atmosphere turbulence for the UTs 
--  and hence the image improvement to be expected from GLAO correction -- have previously been 
estimated by differencing the integrated turbulence measured by DIMM (full atmosphere) and the 
MASS (free atmosphere\hl{above 500~m}) \citep{Sarazin08}.  
This method typically yields substantially larger values of the ground layer fraction than we find 
from the SL-SLODAR data, 
since (i) the median integrated turbulence strength for the free atmosphere from MASS is lower 
than that found from SL-SLODAR (see below), and (ii) from the SLODAR analysis we expect the 
surface layer strength at the height of the UT to be slightly weaker than at the height of the DIMM 
(see section~\ref{sec:ExponentialModel}). \hl{For the SL-SLODAR data, we find a median value for 
the fraction of the total turbulence strength lying above 10~m (UT height) and below 500~m is 0.354. 
Differencing the DIMM and MASS measurements, for the MASS--DIMM data used in this study, yields a median 
ground--layer fraction of 0.636.}

\begin{figure}
 \includegraphics[width=\columnwidth]{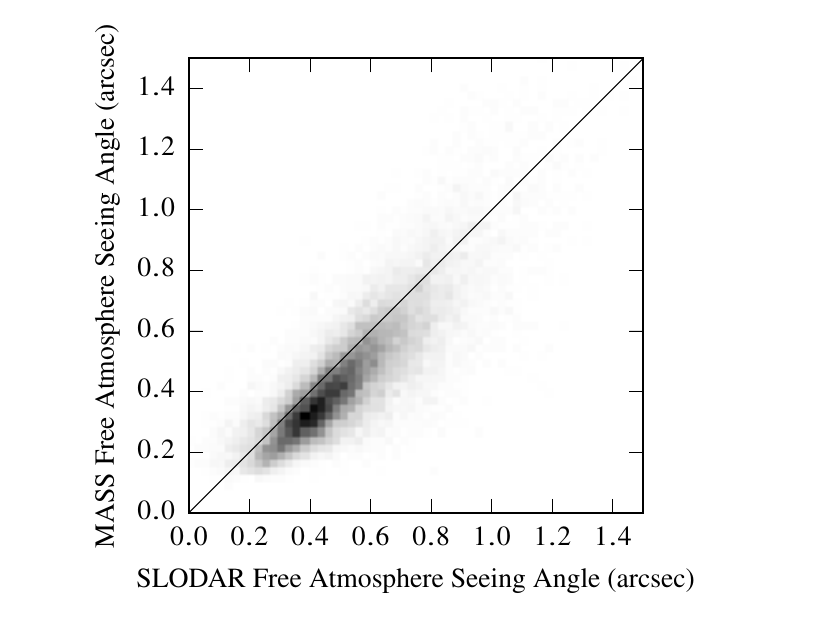}
 \caption{Comparison of contemporaneous SL-SLODAR and MASS measurements of the seeing angle
 for the integrated turbulence above altitude 250~m. The black line shows the $y=x$ case for reference.}
 \label{fig:SLOD_MASS}
\end{figure}

In figure~\ref{fig:SLOD_MASS} we show the comparison of seeing values for contemporaneous 
measurements from the SL-SLODAR and the MASS optical turbulence profiler at Paranal, which is 
coupled with the DIMM monitor on the 7~m tower. 
This comparison comprises 
a total of 291165 contemporaneous measurements on 320 nights between 207 May 23 and 2018 September 19. 
MASS exploits measurements of the scintillation of bright single stars to 
determine the integrated optical turbulence strength in 6 layers, at altitudes 0.5, 1, 2, 4, 8, and 16~km 
above the telescope \citep{Kornilov01}. 
The MASS instrument response function is triangular in the logarithm of altitude, 
for each of these layers. Turbulence below 250~m is not sensed, so that MASS provides a measure 
of the integrated turbulence strength in the `free atmosphere'. 

For comparison with the MASS, we multiply the SL-SLODAR measured profile by the MASS response, 
to find the integrated optical turbulence strength above altitude 250~m. We find that the median 
estimate of the seeing angle for the free--atmosphere from SL-SLODAR (0.507~arcsec) is significantly 
larger than from MASS (0.418~arcsec), for the contemporaneous data, although a strong correlation 
of 0.825 is found between the data sets. The origin of this systematic discrepancy is unknown 
and is currently being investigated, 
\hl{but comparisons between MASS and SCIDAR have previously shown inconsistent results \citep{Masciadri14, Lombardi16, Butterley18}.}

We note that, in this case, relatively small differences in the estimates of the absolute 
turbulence strength for the ground--layer and free-atmosphere produce a large change in the 
estimated {\em fractional} contribution to the turbulence strength from the ground layer. 
For the SL-SLODAR data set we 
find a median surface layer fraction of 37~\%, integrating the turbulence strength 
from the height of the UT observing floor (10~m) to altitude 250~m, relative to the total turbulence above 10~m. 
For the MASS--DIMM data contemporaneous with the SL-SLODAR measurements, a ground--layer fraction of 62~\% is found by differencing the DIMM and MASS. 



Here we have focused on the use of the SL-SLODAR data to model the optimal performance of GLAO 
correction for the VLT, in terms of the reduction of the image FWHM to be expected for correction 
of the optical turbulence up to some altitude above the telescope. We note that SL-SLODAR 
optical turbulence profiles also contain valuable information on the anisoplanatic variations 
of the images to be expected with GLAO correction, and which will form the basis of future studies.

\section{Conclusions}
\label{sec:conc}



The Paranal robotic SL-SLODAR system provides ground layer turbulence profiles up to a maximum altitude 
of 500~m, with 8 resolution elements. 

The instrument produces data in ground wind speeds between 3~m/s and 13~m/s. Above 13~m/s the telescope suffers from too much wind shake. Below 3~m/s performance of the instrument is limited by local turbulence within the instrument enclosure.

The surface layer of turbulence is typically strong, but is generally not resolved by the instrument, so we have fitted an exponential model with a scale height of 5~m to the surface layer to allow the fraction of the surface layer that is below the top of the UT domes to be estimated.

The vertical profile of the ground layer of turbulence is very varied, but in the median case most of the turbulence strength in the ground layer is concentrated within the first 50~m altitude, with relatively weak turbulence at higher altitudes up to 500~m. 

We find good agreement between measurements of the seeing angle from the SL-SLODAR and from the Paranal DIMM seeing monitor, and also for seeing values extracted from the Shack--Hartmann active optics sensor of VLT UT1, adjusting for the height of each instrument above ground level. 

Measurements of free--atmosphere seeing (above 250~m) from the SL-SLODAR are significantly larger than those from the Paranal MASS optical turbulence profiler. 

The SL-SLODAR data suggest that a median improvement in the seeing angle from 0.689~arcsec to 0.481~arcsec at 500~nm would be obtained by fully correcting the ground--layer turbulence between the height of the UTs (taken as 10~m) and altitude 500~m.

\section*{Acknowledgements}

Development of the robotic SL-SLODAR was funded by ESO.
The authors would like to thank the staff at Paranal Observatory for their assistance with the robotic SL-SLODAR project.
TB, JO and RWW are grateful to the Science and Technology Facilities Committee (STFC) for financial support (grant reference ST/P000541/1). 
\hl{The authors would like to thank the anonymous reviewer for their constructive comments that helped to improve the paper.}
This research made use of \textsc{python} including \textsc{numpy}, \textsc{scipy} \citep{numpy} and \textsc{matplotlib} \citep{matplotlib}. 





\bibliographystyle{mnras}
\bibliography{refs} 

\todo{Check refs}




\appendix




\section{Method for SL-SLODAR turbulence profile interpolation with an exponential surface layer model}
\label{sec:appendix}


This section describes the method by which the 8-layer SL-SLODAR profiles with variable resolution are converted into interpolated profiles with fixed resolution.

\subsection{Definitions}

\hl{The atmospheric turbulence profile, unaffected by the response of the instrument, is $C_n^2(h)$.}

The SLODAR profile fitting process involves fitting a model that consists of 8 thin layers of turbulence, labelled $i = 0,1,2,..,7$. These layers are evenly spaced at heights
\begin{equation}
	h_i = i \delta h ,
\end{equation}
where the layer spacing, $\delta h$, is given by
\begin{equation}
	\delta h = X \frac{w}{\theta} .
\end{equation}
Here, $w$ is the subaperture width, $\theta$ is target star separation and $X$ is the airmass, which is given (approximately) by
\begin{equation}
	X = \frac{1}{\cos z} ,
\end{equation}
where $z$ is the zenith angle.

The (idealised) measured profile is given by
\begin{equation}
	J_i = \int_0^\infty C_n^2(h) T_i(h) dh ,
	\label{eq:meas}
\end{equation}
where the triangular `response functions' $T_i(h)$ are given by
\begin{equation}
	T_i(h) =
	\begin{cases}
		0 & \for | h - i \delta h | \geq \delta h \\
		\frac{1}{\delta h} (\delta h - | h - i \delta h | ) & \for | h - i \delta h | < \delta h .
	\end{cases}
	\label{eq:response}
\end{equation}

\begin{figure} 
    \begin{center}
    \includegraphics[width=\columnwidth]{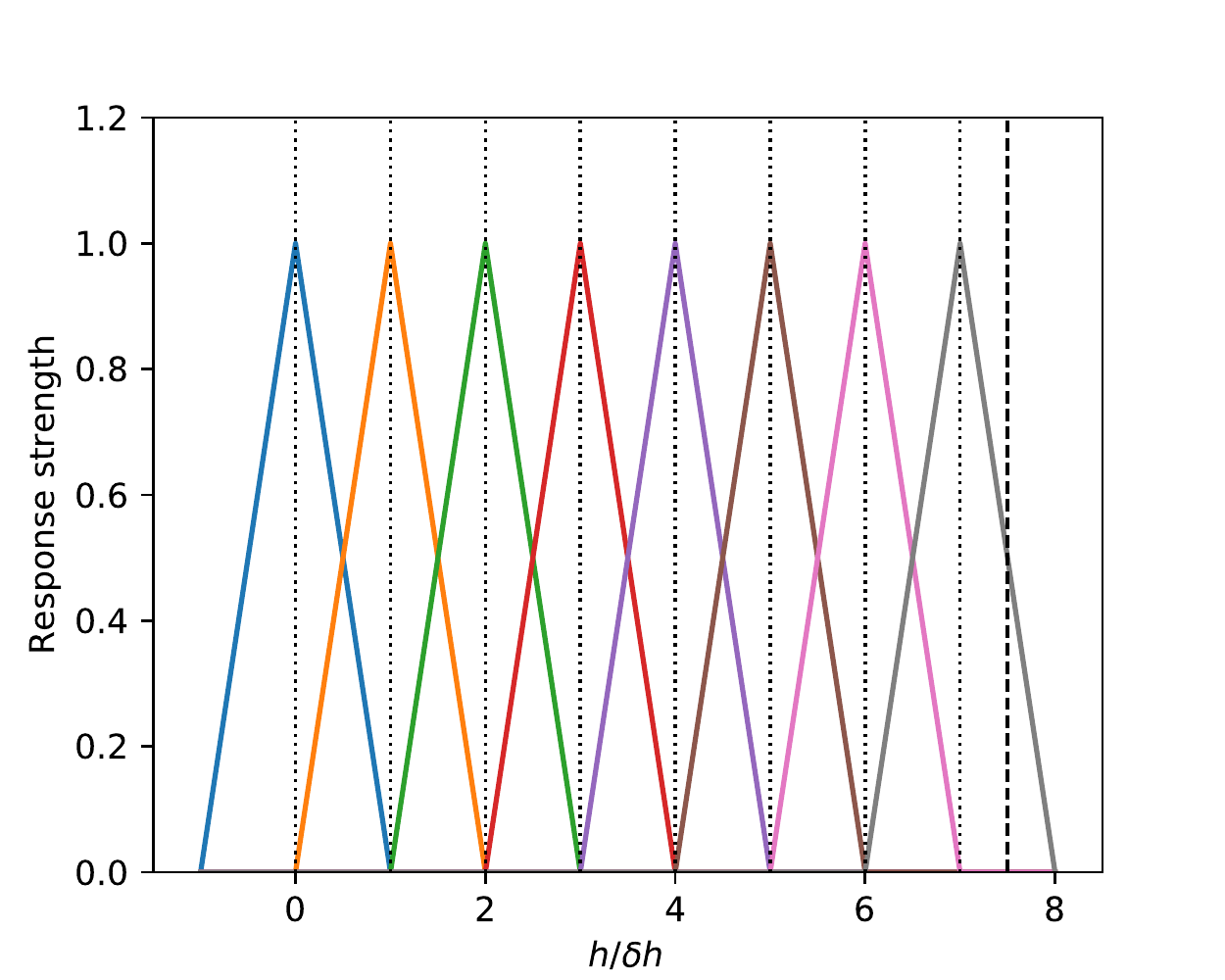}
    \caption{SL-SLODAR response functions. The vertical dotted lines indicate the heights of the 8 reconstructed layers. Only a real layer of turbulence that coincides with one of these layers will appear in a single bin; a layer in between will be split between adjacent reconstructed bins. The vertical broken line shows the cutoff height.}
    \label{fig:response}
    \end{center}
\end{figure}

The response functions\footnote{Here the term `response functions' is used with the same meaning as in the MASS literature. Not to be confused with SL-SLODAR `impulse response functions', which are the reference functions that are fitted to the slope cross-covariance to retrieve the profile.} $T_i(h)$ are shown in Figure~\ref{fig:response}. They show how a layer of turbulence at a given height would be seen by the instrument. For example, a layer at height $1.5 \delta h$ would appear in the reconstructed profile with its strength divided between the model layers at $\delta h$ and $2 \delta h$.

We define the cutoff height, which represents the maximum sensing height of the instrument, to be $h_{\textrm{cutoff}} = 7.5 \delta h$. This is chosen as the height at which the response function of the highest fitted layer drops to 0.5.


\subsection{Exponential surface layer model}

We assume that we can separate some component of the profile into a surface layer model described by an exponential function. We write the model as
\begin{equation}
 	m(h) = J_{\textrm{SL}} n(h) ,
\end{equation}
where $J_{\textrm{SL}}$ is the turbulence strength and $n(h)$ is the normalised exponential model,
\begin{equation}
	n(h) = A \exp \left( \frac{-(h+h_{\textrm{slodar}})}{h_{\textrm{SL}}} \right) ,
\end{equation}
where $h_{\textrm{slodar}}$ is the height of the SL-SLODAR instrument above the ground and $h_{\textrm{SL}}$ is the scale height of the surface layer model. $A$ is a normalisation constant such that
\begin{equation}
	\int_0^\infty n(h) dh = 1.
\end{equation}
We adopt values of  $h_{\textrm{slodar}} = 2$~m and $h_{\textrm{SL}} = 5$~m.

In order to fit this model to the existing 8-profile we first need to map it to the same 8 resolution elements.
The (relative) strengths in each layer of the exponential model are given by
\begin{equation}
 	N_i = \int_0^\infty n(h) T_i(h) dh .
\end{equation}
After numerically evaluating $N_i$, any values $<  0.02$ are set to 0 and the remaining values are then renormalised such that
\begin{equation}
	\sum_{i=0}^{7} N_i = 1 .
\end{equation}

The strength of the surface layer component is determined by finding the maximum possible value of $J_{\textrm{SL}}$ such that $J_{\textrm{SL}} N_i \leq J_i$ for all $i$. In other words, we attribute as much turbulence strength to the exponential component as we can without allowing any of the residual layer strengths to become negative.

The residual 8-layer profile with the exponential surface layer component removed is then given by
\begin{equation}
	J_i^\prime = J_i - J_{\textrm{SL}} N_i .
\end{equation}

\subsection{Interpolation}

The interpolated profile consists of the sum of the exponential surface layer model and the 8 residual layers, each of which is distributed over a range of altitudes defined by its corresponding triangular response function. The interpolated profile is given by
\begin{equation}
	\mathcal{C}_n^2(h) = J_{\textrm{SL}} n(h) + \sum_{i=0}^7 J_i^\prime T_i^\prime (h)
\end{equation}
where $T_i^\prime (h)$ are the response functions defined in equation~\ref{eq:response}, scaled such that
\begin{equation}
	\int_0^\infty T_i^\prime (h) dh = 1 .
\end{equation}
Note that the $i=0$ case has a different normalisation factor because the first response function, $T_0 (h)$, extends outside the integration range (see Figure~\ref{fig:response}). The normalised response functions are 
\begin{equation}
	T_i^\prime (h) = 
	\begin{cases}
		\frac{2}{\delta h} T_i(h) & \for i=0 \\
		\frac{1}{\delta h} T_i(h) & \for i=1,2,3,...,7.
	\end{cases}
\end{equation}

The total $C_n^2 dh$ in the profile is conserved i.e.
\begin{equation}
	\int_0^\infty C_n^2 dh =  \int_0^\infty \mathcal{C}_n^2 dh .
\end{equation}

\subsection{Example}

\begin{figure}
	\includegraphics[width=\columnwidth]{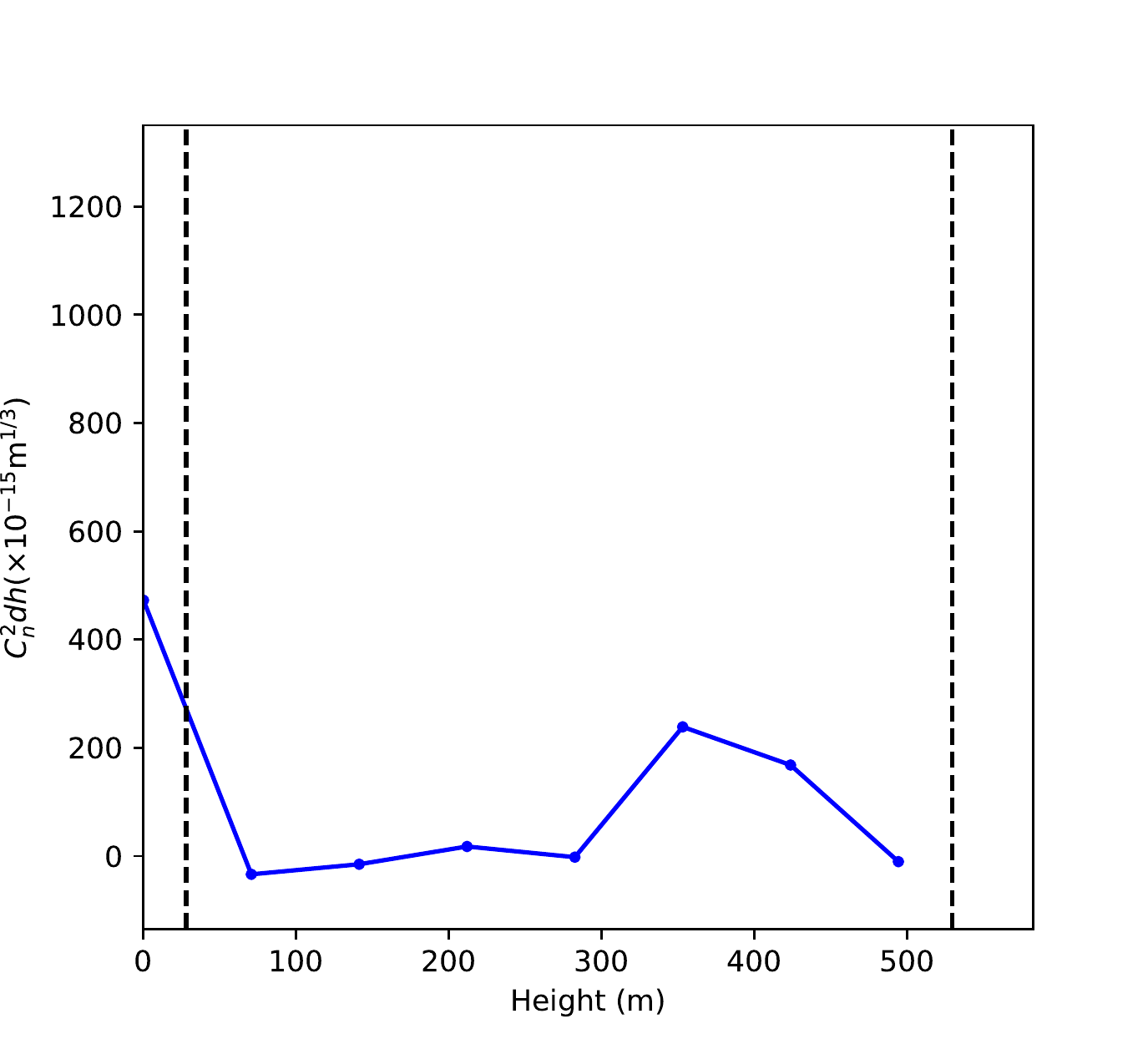}
	\includegraphics[width=\columnwidth]{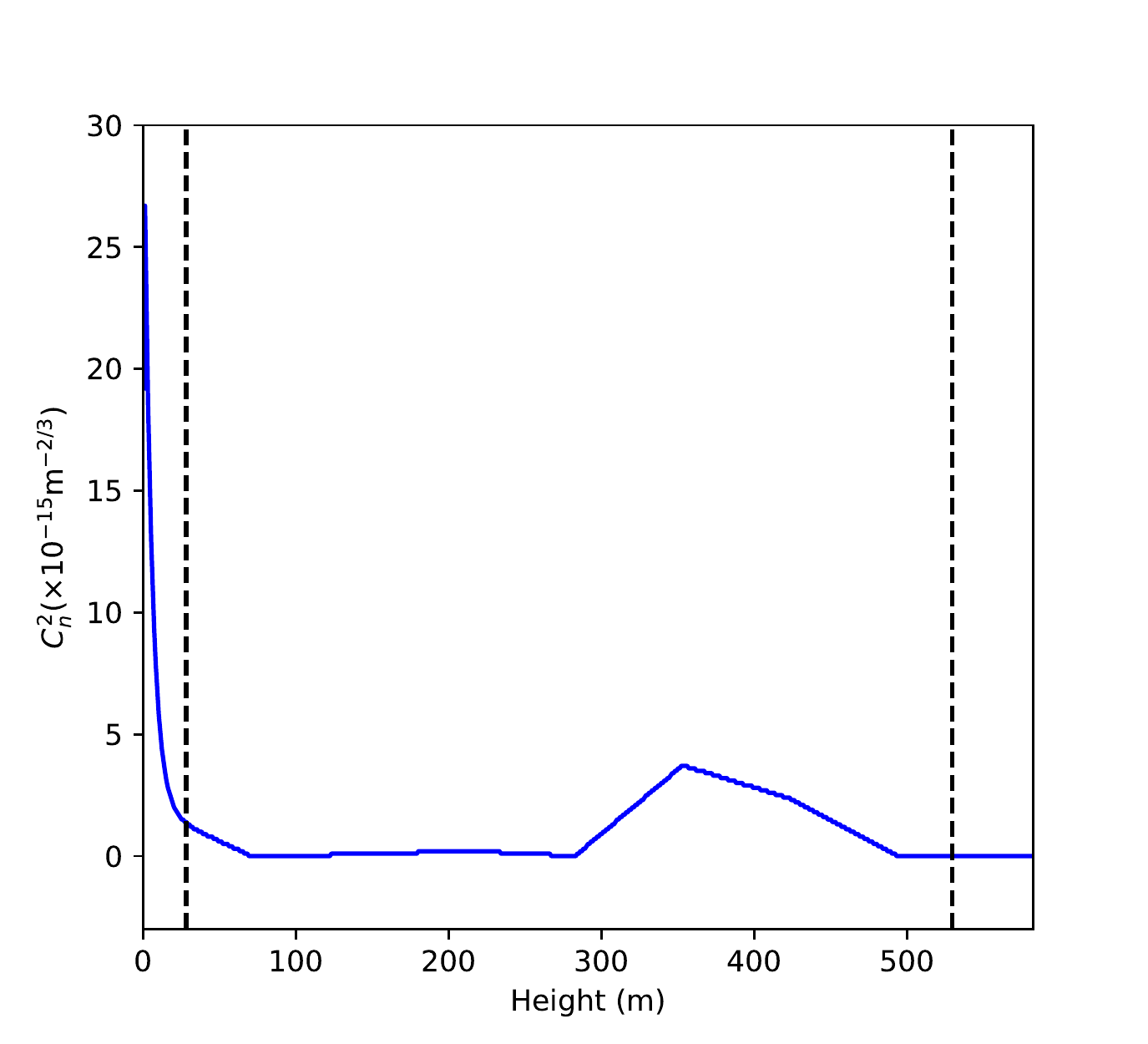}
	\caption{\emph{Top:} Example of a `raw' 8-layer SL-SLODAR profile. \emph{Bottom:} Corrected version of the same profile with samples every 1~m -- the exponential-model surface layer yields a better estimate of the surface layer contribution below the dome height. The broken lines indicate the UT dome height and maximum profiling height.}
    \label{fig:interpExample}
\end{figure}

Figure~\ref{fig:interpExample} shows an example profile before and after interpolation with the inclusion of the exponential surface layer mode. Some of the values from the calculation are shown in table~\ref{tab:interpExample} to help to illustrate the process.

\begin{table}
	\centering
	\caption{\label{tab:interpExample} Example turbulence strength values in the surface layer model calculation. The columns show (i) layer index; (ii) raw profile; (iii) exponential surface layer component; (iv) profile with exponential component subtracted.}
	\begin{tabular}{cccc}
	\hline
	$i$	& $J_i$~($\times 10^{-15}\textrm{m}{^{1/3}}$)	& $J_{\textrm{SL}} N_i$~($\times 10^{-15}\textrm{m}{^{1/3}}$)		& $J_i^\prime$~($\times 10^{-15}\textrm{m}{^{1/3}}$) \\
	\hline \hline
	0	& 270.7	& 142.4	& 128.3 \\
	1	& 10.6	& 10.6	& 0.0 \\
	2	& 4.2		& 0.0		& 4.2 \\
	3	& 13.5	& 0.0		& 13.5 \\
	4	& 0.0		& 0.0		& 0.0 \\
	5	& 230.9	& 0.0		& 230.9 \\
	6	& 146.4	& 0.0		& 146.4 \\
	7	& 0.0		& 0.0		& 0.0 \\
	\hline
	\end{tabular}
\end{table}


\subsection{Limitations}

One should bear the following points in mind when making use of SL-SLODAR profiles that have been interpolated as described above.
\begin{itemize}
\item A fixed scale height is assumed for the exponential model ($h_{\textrm{SL}} = 5$~m). There will be many times when the surface layer does not adhere to this model.
\item The interpolation method has the effect of `blurring' the profile. It is roughly equivalent to convolving the raw 8-layer profile with a triangular function (with a modification at the ground). Note that feeding the interpolated profile back into equation~\ref{eq:meas} will \emph{not} yield the original 8-layer profile.
\end{itemize}



\todototoc
\listoftodos[To do list]


\bsp	
\label{lastpage}
\end{document}